\documentclass[longauth]{aa}  

\usepackage{graphicx}
\usepackage[hidelinks]{hyperref}
\usepackage{txfonts}
\usepackage{xcolor}
\usepackage{enumitem}
\usepackage{cancel}
\hypersetup{colorlinks, linkcolor={blue!70!black}, citecolor={blue!70!black}, urlcolor={blue!70!black}}
\newcommand{\kms}{km\,s$^{-1}$}

\begin{document}

   \title{CAVITY, Calar Alto Void Integral-field Treasury surveY \\and project extension}
   \titlerunning{CAVITY and CAVITY+}

  % \subtitle{I. Overviewing the $\kappa$-mechanism}

\author{I. P\'erez\inst{1,2}
\and 
S. Verley\inst{1,2}
\and
L. S\'anchez-Menguiano\inst{1,2}
\and
T. Ruiz-Lara\inst{1,2}
\and
R. Garc\'ia-Benito\inst{3}
\and
S. Duarte Puertas\inst{1,2,4}
\and
A. Jim\'enez\inst{1}
\and
J. Dom\'inguez-G\'omez\inst{1}
\and
D. Espada\inst{1,2}
\and
R. F. Peletier\inst{5}
\and
J. Rom\'an\inst{6,7}
\and
M. I. Rodr\'iguez\inst{1}
\and
M. Argudo-Fern\'andez\inst{1,2}
\and
G. Torres-R\'ios\inst{1}
\and
B. Bidaran\inst{1}
\and
M. Alc\'azar-Laynez\inst{1}
\and
R. van de Weygaert\inst{5}
\and
S.F. Sánchez\inst{8}
\and
U. Lisenfeld\inst{1,2}
\and
A. Zurita\inst{1,2}
\and
E. Florido\inst{1,2}
\and
J. M. van der Hulst\inst{5}
\and
G. Bl\'azquez-Calero\inst{3}
\and
P. Villalba-Gonz\'alez\inst{9}
\and
I. del Moral-Castro\inst{10}
\and
P. S\'anchez Alarc\'on\inst{6}
\and
A. Lugo-Aranda\inst{11}
\and 
D. Walo-Mart\'in\inst{6}
\and
A. Conrado\inst{3}
\and
R. Gonz\'alez Delgado\inst{3}
\and
J. Falc\'on-Barroso\inst{6,7}
\and
A. Ferr\'e-Mateu\inst{6,7}
\and
M. Hern\'andez-S\'anchez\inst{12}
\and 
P. Awad\inst{5,13}
\and
K. Kreckel\inst{14}
\and
H. Courtois\inst{15}
\and
R. Espada-Miura\inst{1}
\and
M. Rela\~no\inst{1,2}
\and
L. Galbany\inst{16}
\and
P. S\'anchez-Bl\'azquez\inst{17}
\and
E. P\'erez-Montero\inst{3}
\and
M. S\'anchez-Portal\inst{18}
\and
A. Bongiovanni\inst{18} 
\and
S. Planelles\inst{12}
\and
V. Quilis\inst{12}
\and 
M Aubert\inst{15}
\and
D. Guinet\inst{15}
\and
D. Pomar\'ede\inst{15}
\and
A. Weijmans\inst{19}
\and
M. A. Raj\inst{5}
\and 
M. A. Arag\'on-Calvo\inst{8}
\and
M. Azzaro\inst{20}
\and
G. Bergond\inst{20}
\and
M. Blazek\inst{20}
\and
S. Cikota\inst{20}
\and
A. Fern\'andez-Mart\'in\inst{20}
\and
A. Gardini\inst{20}
\and
A. Guijarro\inst{20}
\and
I. Hermelo\inst{20}
\and
P. Mart\'in\inst{20}
\and
J. I. Vico Linares\inst{20}
 }
\institute{Universidad de Granada, Departamento de F\'isica Te\'orica y del Cosmos, Campus Fuentenueva, Edificio Mecenas, E-18071, Granada, Spain. \email{isa@ugr.es}  
\and
Instituto Carlos I de F\'isica Te\'orica y Computacional, Facultad de Ciencias, E-18071 Granada, Spain  
\and
Instituto de Astrof\'isica de Andaluc\'ia - CSIC, Glorieta de la Astronom\'ia s.n., 18008 Granada, Spain  
\and
D\'epartement de Physique, de G\'enie Physique et d’Optique, Universit\'e Laval, and Centre de Recherche en Astrophysique du Qu\'ebec (CRAQ), Québec, QC, G1V 0A6, Canada
\and
Kapteyn Astronomical Institute, University of Groningen, Landleven 12, 9747 AD Groningen, The Netherlands
\and
Instituto de Astrof\'isica de Canarias, V\'ia L\'actea s/n, 38205 La Laguna, Tenerife, Spain
\and
Departamento de Astrof\'isica, Universidad de La Laguna, 38200 La Laguna, Tenerife, Spain
\and
Universidad Nacional Aut\'onoma de M\'exico, Instituto de Astronom\'ia, AP 106, Ensenada 22800, BC, M\'exico
\and
Department of Physics and Astronomy, University of British Columbia, Vancouver, BC V6T 1Z1, Canada
\and
Instituto de Astrof\'isica, Facultad de F\'isica, Pontificia Universidad Cat\'olica de Chile, Av. Vicu\~na Mackenna 4860, Santiago, Chile 
\and
Instituto de Astronom\'ia, Universidad Nacional Aut\'onoma de M\'exico, A. P. 70-264, 04510, M\'exico, D.F., Mexico
\and
Departament d’Astronomia i Astrof\'isica, Universitat de Val\`encia, E-46100 Burjassot (Val\`encia), Spain
\and
Bernoulli Institute for Mathematics, Computer Science and Artificial Intelligence, University of Groningen, 9700 AK, Groningen, The Netherlands
\and
Astronomisches Rechen-Institut, Zentrum f\"ur Astronomie der Universit\"at Heidelberg, 69120 Heidelberg, Germany
\and
Universit\'e Claude Bernard Lyon 1, IUF, IP2I Lyon, 4 rue Enrico Fermi, Villeurbanne, 69622, France
\and
 ICE-CSIC, Carrer de Can Magrans, 08193 Cerdanyola del Vall\'es, Barcelona, Spain
\and
Departamento de F\'isica de la Tierra y Astrof\'isica \& IPARCOS, Universidad Complutense de Madrid, E-28040, Madrid, Spain
\and
Institut de Radioastronomie Millim\'etrique (IRAM), Av. Divina Pastora 7, Local 20, 18012 Granada, Spain
\and
School of Physics and Astronomy, Univ. of St Andrews, North Haugh, St Andrews, KY16 9SS, UK
\and
Centro Astron\'omico Hispano en Andaluc\'ia, Observatorio de Calar Alto, Sierra de los Filabres, 04550 G\'ergal, Almer\'ia, Spain
}

   \date{Received --- / Accepted ---}

% \abstract{}{}{}{}{} 
% 5 {} token are mandatory
 
  \abstract{We have learnt in the last decades that the majority of galaxies belong to high density regions interconnected in a sponge-like fashion. This large-scale structure is characterised by clusters, filaments, and walls, where most galaxies concentrate, but also under-dense regions called voids. The void regions and the galaxies within represent an ideal place for the study of galaxy formation and evolution, as they are largely unaffected by the complex physical processes that transform galaxies in high-density environments. The void galaxies may hold the key to answer current challenges to the $\Lambda$CDM paradigm as well. The CAVITY survey is a Legacy project approved by the Calar Alto Observatory to obtain spatially resolved spectroscopic information of $\sim300$ void galaxies in the Local Universe (0.005 < z < 0.050), covering -17.0 to -21.5 in $\rm r$ band absolute magnitude. It officially started in January 2021 and has been awarded 110 useful dark observing nights at the 3.5 m telescope using the PMAS spectrograph. Complementary follow-up projects, including deep optical imaging, integrated as well as resolved CO data, and integrated HI spectra, have joined the PMAS observations and naturally complete the scientific aim of characterising galaxies in cosmic voids. The extension data has been named CAVITY+. The data will be available to the whole community in different data releases, the first of which is planned for July 2024, and it will provide the community with PMAS datacubes for around 100 void galaxies through a user friendly and well documented database platform. Here, we present the survey, sample selection, data reduction, quality control schemes, science goals, and some examples of the scientific power of the CAVITY and CAVITY+ data.}

   \maketitle
%
%-------------------------------------------------------------------

\section{Introduction}

Galaxies are not uniformly distributed throughout space at large scales, that is, at scales larger than 10~Mpc. Instead, they form a foam-like structure characterised by elongated filaments, sheet-like walls, dense clusters, and under-dense regions in between what are called voids. These voids are an integral part of the cosmic web \cite[see e.g.][for a review]{2011IJMPS...1...41V}, representing around 70\% of the Universe's volume, but they only host around 10\% of the mass \citep{2018MNRAS.473.1195L}. Voids appear from the primordial Gaussian field of density fluctuations, and as a result of their under-density, they represent regions of weaker gravity and therefore grow faster than the Hubble flow. As the voids expand, matter is pushed in between them into the filaments and sheets that contract faster by their own gravity. 

In a seminal work, \citet{2014MNRAS.441.2923C} showed from the analysis of dark-matter-only numerical cosmological simulations, the complex evolution of the Universe's structures. These simulations showed substructures as being part of the hierarchical nature of the cosmic web. This complexity in the large-scale structure poses a challenge when trying to identify the cosmic web patterns. The halo mass distribution function of voids is populated by low-mass objects, as compared to the halo mass distribution of walls, filaments, and clusters. As the Universe evolves, the lowest density regions tend to expand and become even emptier, as matter seems to flow from the low-density regions to the denser environments, and therefore, the dynamical stage of the void can reveal the story of its formation and future fate.
  
Voids therefore represent the key to the arrangement of the large-scale structure of the cosmic web. Despite their name, void structures contain galaxies. These void galaxies, as they evolve in a low-density Universe, {show more recent star formation as compared to galaxies in denser environments} \cite[as suggested by some works, e.g.][]{2012MNRAS.425..641L}, and they provide clues as to how halos assemble their mass in the early Universe and can help us understand the time when internal mechanisms take over in galaxy evolution and how they depend on the large-scale structure. 

We know that galaxies in the Universe broadly come in two general flavours: galaxies that are forming stars,and galaxies that are quiescent. The physical mechanisms that transform star-forming galaxies into quiescent passive systems are still unclear despite the fact that it is a key stage in galaxy evolution. Some mechanisms have been proposed: outflows to remove the gas \cite[][]{2005Natur.433..604D,2006ApJ...652..864H,2012MNRAS.425L..66M}, stripping of the gas \cite[e.g.][]{2000Sci...288.1617Q}, or strangulation \cite[e.g.][]{2006MNRAS.370.1445D}. It is also well established that galaxy interactions can cause an enhancement of the star-formation rate \cite[e.g.][]{1978ApJ...219...46L,2015ApJS..218....6B,2015ApJ...807L..16K}, the largest enhancement occurring in equal-mass mergers after the first pericentre passage and, much stronger, during coalescence \cite[e.g.][]{2004MNRAS.355..874N,2012MNRAS.426..549S}.
Void galaxies reside in environments that are largely unaffected by the processes modifying galaxies in denser environments, and this should be reflected in their star-formation histories (SFHs). By removing the physical processes of dense environments, we are left to study a `simplified', pristine environment that can help us understand the main drivers of galaxy transformation. Observations seem to indicate that void galaxies are similar to galaxies in denser regions, but they tend to be bluer and of later type morphologies than galaxies in denser environments \cite[e.g.][]{2004ApJ...617...50R,2005ApJ...624..571R,2007ApJ...658..898P,2008ApJ...673..715C}. Some works show indications that the specific star-formation rate (sSFR = SFR/M$_{\star}$) of void galaxies might be slightly enhanced with respect to non-void galaxies \cite[][]{2017MNRAS.464..666B}, but the works are inconclusive.   Recent results \citep{2022A&A...658A.124D} for a sample of 20 void galaxies hint towards some differences between the star-formation efficiency of void and non-void galaxies, excluding those in clusters. They also showed that the molecular content seems to be similar, although a larger sample is necessary to confirm the trends. 

It is increasingly becoming clear that the environment has an influence on the stellar mass of galaxies, as void galaxies do not seem to follow the morphology-density relation \cite[][]{1980ApJ...236..351D}, as they lack very low-mass galaxies, $\rm M_{\star} < 10^{7}M_{\odot}$ \cite[][]{2011ASSP...27...17V,2016MNRAS.458..394B}. There are discrepancies about the present metallicity content of the gas in void galaxies. Some studies show systematically lower metallicities \cite[][]{2011MNRAS.415.1188P}. However, a study of luminous void galaxies showed no deviation from the standard mass-metallicity relation \cite[][]{2015ApJ...798L..15K}. With integral field spectroscopy (IFS) data, we can directly measure the metallicity gradient and determine the metallicity at a fixed galaxy effective radius. Furthermore, given the evidence for ongoing cold accretion of pristine material in some void galaxies \cite[][]{2009ApJ...696L...6S}, this could be identifiable as asymmetries in 2D metallicity maps.   

Observationally, the controversy remains as to whether or not the properties, including the mass of the dark halo, of galaxies in voids differ from comparable objects in walls and filaments. Some of the controversy could be due to a lack of a proper characterisation of the void substructure and the void dynamical stage. Furthermore, all of these studies lack a view on how star formation occurred in the past and when and how the mass was put together. 
The SFHs, including the run of star formation with time and the cosmic chemical enrichment of stars, hold the key to understanding the baryonic mass assembly of galaxies in these low-density environments. What is observed in galaxies is the accumulated history of multiple generations of stars. The properties of the stars currently located in a galaxy reflect the different episodes of star formation and the interstellar medium enrichment processes undergone. Therefore, by deriving the current stellar composition in galaxies, we can trace back the history of how stars have been formed over the entire evolution of the galaxy. In other words, we can obtain its SFH.
Indeed, in a recent study carried out on the Sloan Digital Sky Survey (SDSS) spectra of a well-defined sample of void, wall, filament, and cluster galaxies, defined as the CAVITY parent sample (see Sect.~\ref{sample}), we concluded that the stellar mass assembly of void galaxies occurs slower than in denser environments \citep{2023Natur.619..269D}. When analysing the stellar metallicities of the same sample, \citet[][]{2023arXiv231011412D} found that, on average, void galaxies have lower stellar metallicities than galaxies in denser environments.

It has previously been shown that spatially resolved SFHs of galaxies obtained from IFS data can determine the epoch of galaxy structure formation \cite[e.g.][]{2017A&A...608A..27G,2017MNRAS.470L.122P}. In addition, previous studies have demonstrated that deriving SFHs from integrated spectra using full spectral fitting techniques provides robust results as compared to resolved stellar populations \cite[e.g.][]{2012MNRAS.423..406G, 2015A&A...583A..60R,2018A&A...617A..18R}. 
All of these facts made the time ripe for a detailed and spatially resolved study of baryonic mass assembly in  the most under-dense regions of the cosmic web, motivating the development of the CAVITY project, presented in this work. 

In Sect.~\ref{summary}, we present an outline of the CAVITY project. Sect~\ref{sample} describes the void and galaxy selection that conforms the CAVITY parent sample. In Sect.~\ref{pmas_data_reduction}, we detail the on-going observation strategy and data reduction pipeline. Sect.~\ref{QC} outlines the quality assessment performed on the IFS data. In Sect.~\ref{cavityplus}, we describe in detail the project extension, named CAVITY+. We explain the generalities of the CO, HI, and deep optical imaging data. In Sect.~\ref{science}, we present the main project goals and some examples of the scientific power of the CAVITY and CAVITY+ data.

\section{CAVITY project summary}
\label{summary}
Calar Alto Void Integral-field Treasury surveY (CAVITY) is a survey aimed to study galaxies in voids using IFS data. The project was approved in 2020 by the Calar Alto Observatory (CAHA, Spain) as a Legacy project, being awarded 110 useful observing nights at the 3.5 m telescope using the PMAS spectrograph. The observatory officially started scheduling CAVITY nights in January 2021. 

CAVITY is currently addressing in detail the previously discussed issues on SFHs, and the physical properties of the ionised gas. Furthermore, the dark mass content of these galaxies is also of high interest for the project. Since the relation between the baryonic mass and its dark halo is not linear \cite[e.g.][]{2013MNRAS.428.3121M}, in order to assess the whole galaxy assembling we need to first determine the dark matter contribution to the whole potential. We will be able to address the dark content of void galaxies with the IFS data. Observationally, determining the dark mass of void galaxies has been attempted on a sample of void galaxies present in the MaNGA catalogue \cite[][]{2019ApJ...886..153D} finding no significant difference between galaxies located in voids and in denser environments. The CAVITY sample (see Sect.~\ref{sample}) has been specifically designed to study galaxies within voids, as well as sampling voids of different sizes, and at different dynamical stages.
The CAVITY project will also address the predictions of the current cosmological model, the $\Lambda$CDM  model, for the mass assembling of galaxies in different environments. The $\Lambda$CDM  model, which is the parametrisation of the Big-Bang theory with dark matter and dark energy, has been very successful at predicting the main observables in the Universe (large-scale structure, cosmic microwave background, acceleration of the expansion of the Universe, and the abundances of elements). However, it still fails at predicting the scales at which baryon physics dominates, i.e. galaxy scale \cite[e.g.][]{2007ApJ...670..313S,2009MNRAS.398.1027K}. The details of the time-scales at which baryons assembled to form galaxies in the early Universe, and their relation with their hosting dark halo still remains controversial as the baryon physics is difficult to model. CAVITY will serve as the perfect benchmark to confront the state-of-the art numerical predictions from $\Lambda$CDM, which will help us interpret the general context of voids in the formation and evolution of galaxies. 

\begin{figure*}
\centering 
\includegraphics[width = 0.95
\textwidth]{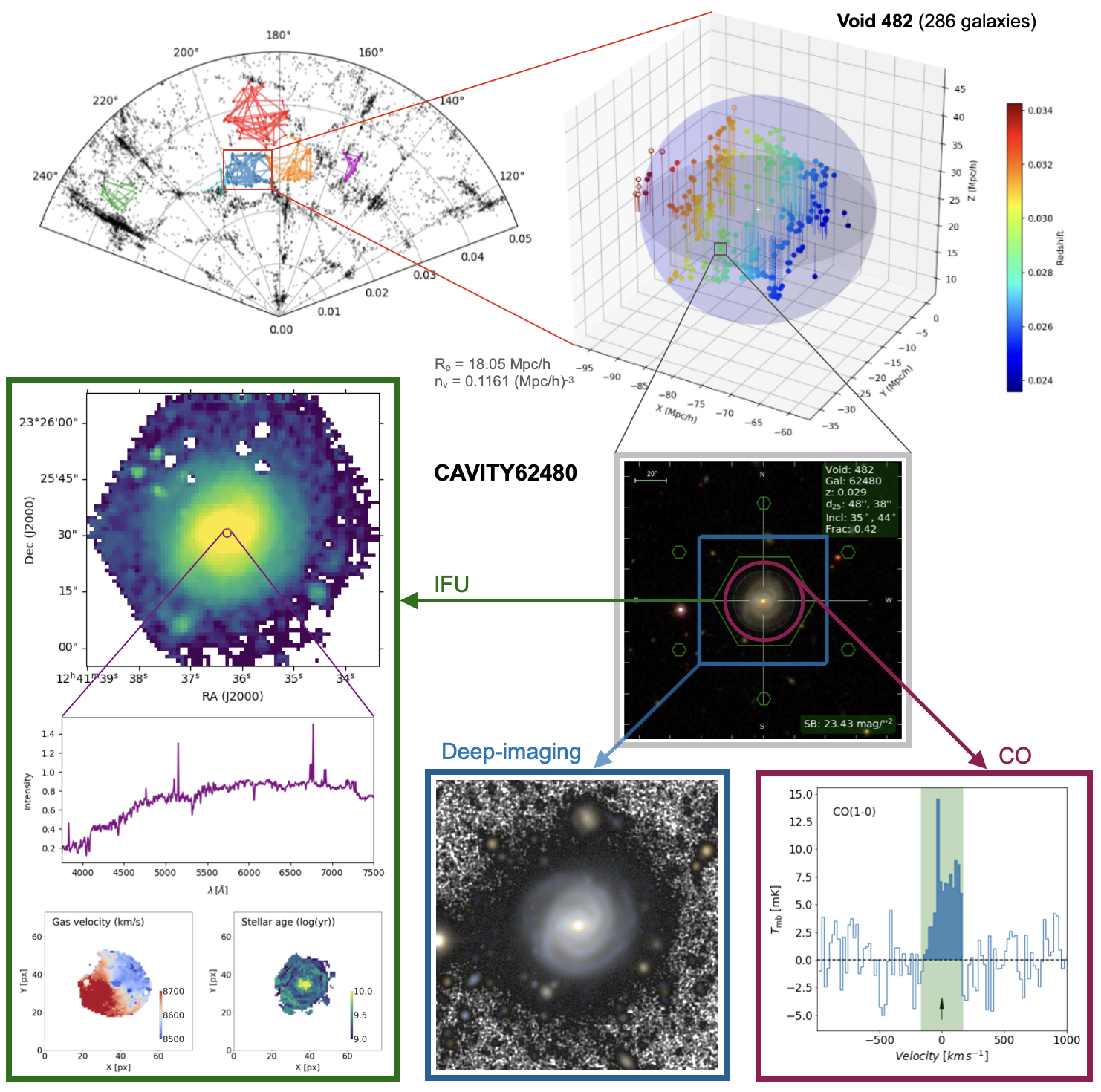} \\   
\caption{Summary figure of CAVITY and project extension. This figure is divided into three main sections. At the top we illustrate some randomly selected CAVITY voids within the large-scale structure of the Universe (as a wedge plot on the left, \cite{2017PASP..129e8005A}) and show the spatial distribution of galaxies within a typical void (void 486; each dot corresponds to a different galaxy colour-coded by redshift). The webs of colour lines link galaxy members and depict the location of each highlighted void. The middle right panel (framed by a grey square) zooms in on CAVITY62480, an example galaxy contained in void 482, showing its SDSS colour image with the PMAS footprint (green hexagon), INT cutout (blue square), and IRAM beam (purple circle) on top of it. The next main section (framed by the green rectangle) is devoted to illustrate the IFS data. We represent the integrated light of the galaxy within the covered wavelength range of the instrument (on top), the spectrum of the central spaxel (in the middle), and the gas velocity and stellar age maps (at the bottom) as examples of the potential of IFS data in deriving spatially resolved distributions of the galaxy properties. The third and final section is devoted to CAVITY+, the complementary observational campaigns that shape the CAVITY project extension. Here we show a coloured image using the INT g- and r-band deep imaging (framed by the blue rectangle, bottom middle panel) and the integrated CO(1-0) spectrum from the IRAM observations (purple rectangle, bottom right panel) for CAVITY62480.}
\label{fig:CAVITYplot} 
\end{figure*}

 Summarising, CAVITY is a spatially resolved survey of void galaxies aiming to: 1) determine how the large-scale environment has influenced the mass assembly of void galaxies; 2) establish how galaxy formation and its properties are dependent on the large-scale environment; 3) find the main drivers for galaxy transformation; and 4) determine the dark matter content of void galaxies. Figure~\ref{fig:CAVITYplot} illustrates the CAVITY project and its unparalleled power to characterise galaxies in voids.

 The IFS data for the first 100 galaxies will be made completely available to the public as part of the first CAVITY Data Release (DR1) through an easy access database planned for July 2024. The full sample and high-level data products are planned to be released by 2025/2026.

\section{Sample selection and characterisation}
\label{sample}
 
 A broad definition of a cosmic void as a large under-dense (average density contrast of about $-0.9$) structure of diameters generally between 20 to 100\,Mpc is a clearly accepted one \cite[e.g.][]{2008MNRAS.387..128P}. However, when it comes to the details of determining the shape of a void and its limits within the large-scale structure, in both galaxy surveys and simulations, it becomes a tricky task that is subject to different detection algorithms that present clear differences among each other but they agree in the generalities \cite[e.g.][]{2008MNRAS.387..933C,2018MNRAS.473.1195L}.
Overall, most void finder strategies in survey data use a conversion from redshift to 3D distance and then estimate a local density field from either the raw or smoothed data, assuming or not some shape priors. Many void finders assume approximately spherical voids \citep{2002ApJ...566..641H}, or sum of spheres or `holes' \cite[e.g.][]{2004ApJ...607..751H,2012MNRAS.421..926P}, while others base their catalogues in void-finder codes that make no assumption on the shape of voids, using some tessellation technique to obtain the density field and a `watershed' threshold \citep[e.g.][]{2014MNRAS.440.1248N,2015A&C.....9....1S}. These void identifications have to take into account that some voids may also show internal substructure known as `tendrils' \citep{1997ApJ...491..421E, 2002ApJ...566..641H, 2004ApJ...607..751H}. 

Once cosmic voids have been identified, one can start identifying galaxies inhabiting these voids. In the nearby Universe ($\rm z<0.1$), several void catalogues have been produced from the SDSS data \citep{2012MNRAS.421..926P, 2014MNRAS.440.1248N, 2022arXiv220201226D}. For the selection of CAVITY galaxies we opted to use the \citet[][]{2012MNRAS.421..926P} catalogue based on SDSS data, which is a well characterised catalogue of nearby voids.
We restricted ourselves to voids in the catalogue {where the full volume of the void is included in the SDSS footprint} within the redshift range 0.005 and 0.050; these redshift limits also allowed us the obtain a full {volume} coverage of voids of large sizes. In addition, we have selected galaxies with intermediate inclinations{, with a final inclination distribution between 20 and 70 degrees,} visually discarding {a few }very edge- and face-on systems{, where the intermediate inclinations derived from LEDA\footnote{http://leda.univ-lyon1.fr/} \citep{2014A&A...570A..13M} and SDSS\footnote{https://classic.sdss.org/dr7/} \citep{2009ApJS..182..543A} where clearly wrongly estimated}. From the initial 79947 galaxies in the sample of 1055 voids in \citet[][]{2012MNRAS.421..926P}, we are left with 19857 in 96 voids that fulfil the criteria of the redshift range and {fitting the full void} within the SDSS footprint. To accurately characterise voids, we establish a criterion requiring at least 20 galaxies per void. This results in a total of 19,732 galaxies distributed across 80 voids. After removing the voids at the edges of the SDSS footprint, {where possibly part of the void volume is missing,} we are left with 8690 galaxies in 42 voids. From these 42 voids we have selected 15 representative voids to cover the full ranges in void effective radius, volume number densities, number of galaxies within voids, and right ascension, to maximise the observability from the Calar Alto Observatory. The selection also ensures that the distribution of some properties of the galaxies within the 15 voids are representative of those within the 42 voids, such as redshift, effective radius fraction, $R_{90}$ Petrosian, $\rm R$ absolute magnitude, $\rm g-r$ colour distributions. This later selection leaves us with a sample of 4866 galaxies, which is considered the CAVITY parent sample. From these, we have removed those objects in common with the cluster catalogue of \citet[][]{2017A&A...602A.100T} to ensure that we had no galaxies in the sample belonging to clusters.

As previously discussed, the membership of a particular galaxy to a void structure or to a higher density environment is a decision dependent on the void detection algorithm used. This is more evident at the edges of the structures. To ensure that the final selection is basically dominated by void galaxies we have imposed an extra selection criterion based on the distance of the galaxy to the centre of the void, characterised by the effective radius parameter ($\rm R_{eff}$) that corresponds to the radius of a spherical void of equal volume. This is a simplification since we know that the intrinsic shape of voids is more complex than a simple spherical representation \citep{2008MNRAS.386.2101N, 2015A&C.....9....1S}. A galaxy needs to be within a radius of 0.8 $\times~\rm R_{eff}$, with the $\rm R_{eff}$ as defined above, to be included in the sample.

\begin{figure}[h]
    \centering 
        \includegraphics[width=\linewidth]{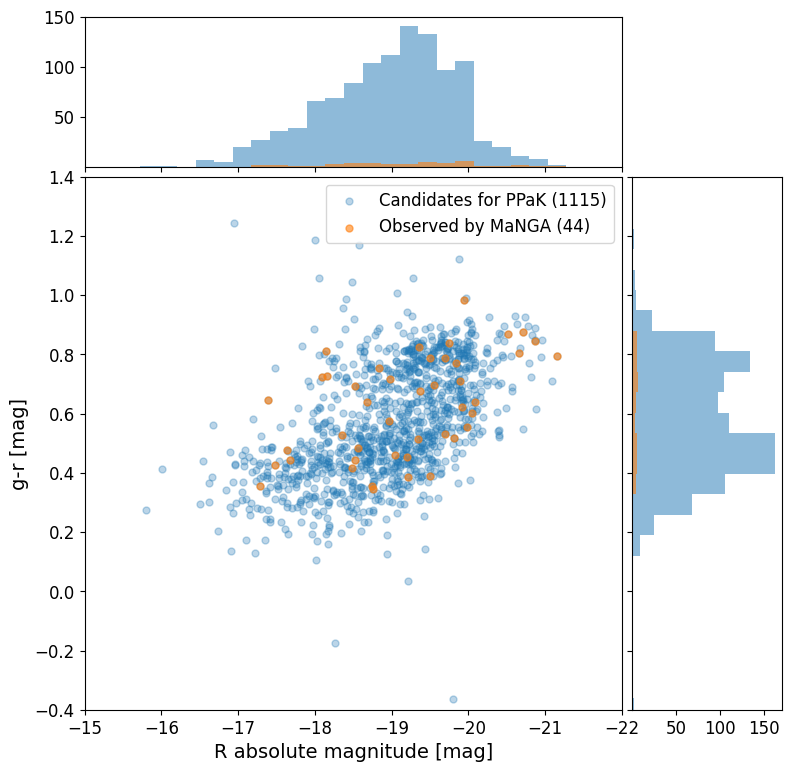}
        \includegraphics[width=\linewidth]{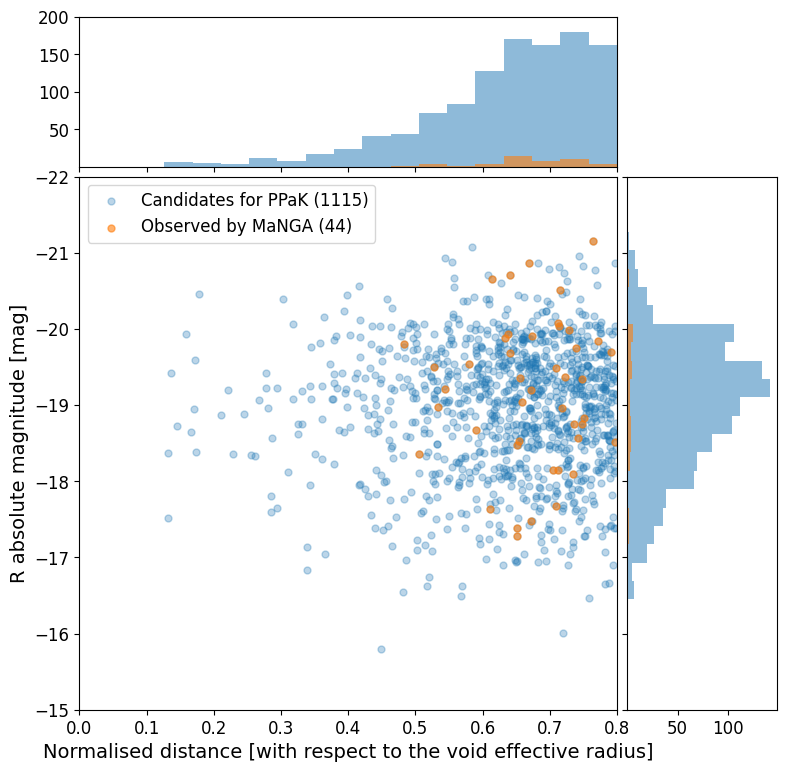}  
    \caption{CAVITY sample characterisation. Top panel: Colour-magnitude diagram of the CAVITY 1115 observable galaxies. Bottom panel: Distribution of the projected distance of the 1115 observable galaxies to the centre of the void normalised to the ${\rm R_{eff}}$ of the void with respect to the {\rm R-band} absolute magnitude. In both panels the 44 MaNGA galaxies included in the parent sample have been indicated in orange.}
    \label{fig:reff_parent}
\end{figure}

{During the sample selection phase we developed a hybrid approach mixing information from well-known catalogues (LEDA and SDSS) as well as a visual inspection (using GALAssify, Alcázar-Laynez et al., in prep.) to revise the sample and discard galaxies that have too bright stars in the PMAS field of view (FoV), very edge-on systems (as described before), and galaxies that show too low surface brightness (SB) to be observed with PMAS (below an average $\mu_{r}$ of 25 mag arcsec$^2$). After this process, we are left with 1115 observable galaxies.} From these 1115 galaxies, 44 of them are included in the MaNGA sample \citep{2015ApJ...798....7B} and one galaxy is in common with the CALIFA survey \citep{2012A&A...538A...8S}. The top and bottom panels of Fig.~\ref{fig:reff_parent} show the Colour-Magnitude diagram and the distribution of center-void distances within the void (normalised to the void $\rm R_{eff}$) of the final observable sample, respectively. 
%-------------------------------------- Two column figure (place early!)

\section{Observation strategy and data reduction}\label{pmas_data_reduction}
 
Observations are performed with the Postdam Multi Aperture Spectrograph \citep[PMAS,][]{2005PASP..117..620R} in the 
PPaK mode \citep{2006PASP..118..129K}. The PPaK fibre bundle consists of 382 fibres of 2.68 arcsec diameter each.
The science bundle is comprised of 331 fibres in an hexagonal configuration covering a FoV of  
74\arcsec $\times$64\arcsec, with a filling factor of $\sim60\%$. Six independent fibre bundles of 6 fibres each 
conform the sky background sampling. These bundles are distributed along a circle at $\sim$72 arcsec from the 
centre of the instrument FoV \citep[see Fig. 5 of][for an overall layout of the PPaK IFS]{2006PASP..118..129K}. 
Finally, there are 15 extra calibration fibres that are not part of the IFS that can be illuminated with 
the light from the spectral-line lamps during science exposures.

Observations are carried out using the V500 grating. This grating has a nominal resolution of 
$\lambda$/$\Delta\lambda$ $\approx$ 850 at $\sim$5000 \AA\ with a full width at half maximum (FWHM) of about 6 \AA, 
covering from 3745 to 7300 \AA\ and thus including from [OII]$\lambda$3727 to 
[SII]$\lambda$6731 in rest-frame of all galaxies of the survey. 
The brightest and largest objects of the sample, around 50\%, will also be observed using grating V1200. This grating has a nominal resolution of 
$\lambda$/$\Delta\lambda$ $\approx$ 1650 at $\sim$4500 \AA\ with a FWHM of about 2.7 \AA, 
covering from 3400 \AA\ to 4750 \AA. The optical system was designed to tolerate some vignetting at the edges of the FoV of the detector. The effect of this vignetting decreases gradually the efficiency at the corners of the CCD involving 30\% of the fibres (15\% of the fibres for the most severe cases), and therefore the wavelength coverage is reduced up to a $\sim$25 \%. The vignetting presents a ring-like structure around the centre of the FoV \citep[see Fig. 11 of][]{2013A&A...549A..87H}. 
The useful wavelength range varies in those fibres affected by vignetting; in the worst cases, it is reduced to 4240-7140 \AA\ and 3650-4650 \AA\ for the V500 and V1200, respectively.

Similarly to the observations carried out for the CALIFA survey, we follow a three-position dithering pattern in order to have a 100\% filling factor on the aperture \citep[given the parallelism between CALIFA and CAVITY observing strategies, we encourage the reader to check][for a complete summary of technical details regarding PMAS data]{2012A&A...538A...8S}. Exposure times for the V500 vary between 1.5 and 3.0 hours of total integration, 
depending on the brightness of the galaxy. For each dithering position, two exposures of 900s are taken for the brightest targets, while for the faintest galaxies four 900s exposures are 
observed. For the 1.5 hour observations, one set of calibration files are taken for galaxy. 
For longer total integration time, two sets of calibration exposures are obtained. 
All observations are gathered at airmasses below 1.4.

The CAVITY pipeline follows the basic reduction procedures and steps of the CALIFA survey 
\citep{2016A&A...594A..36S}. With a new architecture and improvements in efficiency, it fits the 
particularities of the CAVITY survey. It runs in Python 3 and uses Cython code 
\citep{behnel2011cython} for some specific time-consuming tasks. The reduction process includes the propagation of the Poisson plus read-out noise, cosmic rays, bad CCD columns, and 
the effect of vignetting. First, the four different FITS files of the 4k$\times$4k E2V 
detector are combined into a single frame for each exposure. Each frame is cleaned 
of bad pixels, including cosmic rays using \texttt{PyCosmic} 
\citep{2012A&A...545A.137H}. Then, the locations of the spectra on the CCD are measured 
on the continuum lamp and the small instrument flexure offsets are corrected for 
($\leq$ 0.5 pixel). Straylight contribution is measured using the few gaps between 
fibre traces and a 2D straylight map is subtracted to all calibration and science 
frames. After the extraction of all spectra using the traces and the measured FWHM 
of each fibre, the wavelength calibration solution is obtained from the HeHgCd 
arc-lamp exposures, taking into account the previous estimated offsets. The 
spectra are resampled to a linear grid and homogenised to a common spectral 
resolution. The wavelength sampling is 6 \AA\ for the V500 and 0.7 \AA\ for the 
V1200, while the spectral resolution (FWHM) over the entire wavelength range is 
6 \AA\, and 2.3 \AA, respectively. Fibre-to-fibre transmission differences are 
corrected with sky exposures taken during the twilight. Flux calibration is 
applied using the master sensitivity curve derived from the observation of 
standard stars in several photometric nights. The atmospheric extinction 
along wavelength \citep{2007PASP..119.1186S} is corrected taking into account the 
information of the monitored $V$ band extinction by the Calar Alto 
Visual EXtinction monitor (CAVEX). The sky subtraction is performed using 
the 36 sky fibres distributed around the FoV. The 36 spectra, 
taken at the same time as the science fibres, are combined, removing outliers 
using a sigma-clipping rejection algorithm. The average sky spectrum is then 
subtracted to the science frames. The final step consists in spatially 
re-arranging the spectra to create a datacube. The individual frames at 
each dithering pointing are combined into a single frame and a flux-conserving 
inverse-distance weighting scheme \citep{2012A&A...538A...8S} is used to reconstruct 
a spatial image of the final 993 science fibres. The cube is resampled to 
a pixel scale of 1\arcsec/pixel. After the cube reconstruction, 
the effect of differential atmospheric refraction (DAR) is corrected by 
measuring at each wavelength the spatial position of the centroid of the 
galaxy. Finally, the datacube is corrected for Galactic extinction. 
In next versions of the pipeline we foresee an absolute flux calibration 
anchored to SDSS DR16 \citep{2020ApJS..249....3A}.

\section{Data quality}
\label{QC}

In order to ensure that the data comply with the standard requirements for their use by the scientific community, we perform a quality assessment for the data. This is done in a two-step process. First, during the data reduction, the pipeline performs a set of automatic tests on the individual spectra coming from the fibres that are run to identify possible issues with the data reduction. In a later stage, a visual inspection and preliminary analysis of each final reduced datacube are performed by at least two different members of the team to estimate the quality of the data and identify any additional problem with the reduction. In this section we briefly describe both steps, including a description of the quality estimators developed for the final stages of the data reduction pipeline. A thorough assessment of the quality of the data will be presented along with the DR1. 

\subsection{Tests on individual spectra}

As an integral component of the data reduction process, the pipeline integrates a comprehensive array of automated tests, both at the row-stacked spectra and cube level, that are systematically executed and  documented. These tests culminate in the creation of tables and figures that are subsequently compiled into dedicated web-based reports for each night of observation. These serve as valuable tools to assess the data quality and detect potential issues arising from the reduction process. 

In order to verify the accuracy of the wavelength calibration, the nominal and recovered wavelengths of well-known night-sky emission lines in each spectrum are compared before the sky subtraction step. To estimate the central wavelength of these lines, a Gaussian function is fitted to each one, resulting in 331 estimations of the relative offsets between the nominal and recovered wavelengths. The median value derived from this is approximately 0.1 \AA.

The spectral resolution of a dataset is influenced by various factors, including internal focus issues, tracing/extraction errors, and wavelength calibration inaccuracies. To estimate the actual spectral resolution for each science frame, we fit the strongest night-sky emission lines using a Gaussian function before subtracting the night-sky spectrum. The FWHM values of these lines provide reliable estimates of the final resolution at their respective wavelengths. The empirical resolution for the V500 derived from night-sky lines is $~\sim$6.5 \AA. {This corresponds to a velocity resolution} of $\sigma$ $\sim$ 150 km/s.

To assess instrumental dispersion, arc-lamp frames can be used instead of science frames. However, the arc-lamp exposures are shorter and not affected by resolution degradation caused by spectral drift. The derived instrumental dispersions from arc-lamp frames are smaller and more precise. Nevertheless, the final resolution achieved in the data is closer to that estimated from night-sky lines than from arc lamps.

 The galaxies in CAVITY are chosen to fit within the FoV of the central bundle of fibres, ensuring that most of the 36 sky fibres remain uncontaminated by galaxy light. The subtraction process is relatively straightforward, involving the measurement of flux from prominent night-sky lines before and after subtracting the estimated night-sky spectrum.

\subsection{Inspection of datacubes}
The quality control of the datacubes is carried out based on a series of graphics that are automatically generated for each galaxy after one round of observations is reduced. These graphics include: i) a white image (integrated light from 4500 to 7000 \AA) of the galaxy together with a map showing its negative intensity values; ii) the spectrum of the centre of the galaxy; iii) a two-dimensional map together with an histogram of the continuum S/N; iv) maps showing two different binning schemes common in stellar population studies (Voronoi tessellation presented in \citealt{2003MNRAS.342..345C}, and a typical elliptical binning) together with the radial profile of the S/N in the case of the Voronoi bins; v) S/N maps and histograms for the H$\alpha$ and H$\beta$ emission lines.

All these plots facilitate the execution of the quality control and help highlight problems with the instrument or those resulting from the data reduction, as well as identify low S/N data, targets that are unsuitable for analysis, or pointings observed under non-optimal observing conditions (affecting the cube reconstruction and then to be repeated). The inclusion of the Voronoi binning map gives an idea of the suitability of each galaxy to perform spatially resolved studies of stellar population properties and stellar kinematics. In addition, the datacubes themselves are inspected to evaluate the presence of any evident problem that might affect the shape of the spectra.

In order to inform of the outcome of the data quality assessment, each reviewer has to fill a report for each galaxy summarising the results of the evaluation. We defined four flags that identify the most problematic and/or common issues, namely: i) CUBE: if the overall reconstruction of the cube is wrong; ii) SPECTRUM: if they present any distortion or odd feature; iii) SPATIAL$\_$BINNING$\_$CENTRE: if the centre used in the adopted binning schemes, chosen as the bright spaxel in the cube, is incorrect; and iv) VORONOI: if the number of resulting Voronoi bins is low, or the sizes are very large to properly study the spatially resolved properties of the stellar populations. Besides, any other spotted issue not covered by these flags can be reported as additional comments.

The most common issues reported so far during the IFS quality control are
for instance the presence of several dead fibres in the north-west part of the FoV, a wrong sky emission subtraction resulting in a north-south gradient, or the presence of some sky lines residuals in the spectra not properly flagged as bad pixels in the datacube. None of these issues are irreversible and some procedures can be applied to correct these effects (this will further discussed in DR1 paper). %Of the XX galaxies reduced up to date, we only report $\sim 7\%$ of cases where the cube reconstruction was unsuccessful and the datacubes are not suitable for science. %{\color{olive} TRL: Re-observations of particular pointings sufficient in some cases to solve the problem.}

%a) A white image (integrated light from 4500 to 7000 \AA) of the galaxy together with a map showing negative intensity values. The latter can help to highlight any pattern that resulting from a problem with the reduction pipeline. For example, the presence of dead fibres clearly stand out in this figures, as well as a wrong sky emission subtraction or any problem with the cube reconstruction; b) The spectrum of the centre of the galaxy, which facilitates for instance the observation of any issue related with the CCD. For example, the appearance of a jump in the spectra of some galaxies at around 5750 \AA\, exposed a problem it arose with one of the detectors (with was afterwards solved), which sensibility suffered a drop that was not accounted by the reduction pipeline. 
\subsection{Propagated noise assessment}

An important part of the data reduction strategy  described in Sect.~\ref{pmas_data_reduction} involves the propagation of spectroscopic errors in the datacubes. In order to assess the quality of this complex and essential step, and following \citet{2013A&A...549A..87H}, we thoroughly compare the propagated errors of the datacubes with the residuals from full spectral fitting techniques assuming that such residuals are mainly linked to noise. In particular, we make use of the combination of well-tested codes such as {\tt pPXF} and {\tt GANDALF} \citep[][]{2004PASP..116..138C, 2006MNRAS.366.1151S, 2006MNRAS.369..529F}, extensively used in the characterisation of the properties and kinematics of stellar populations in galaxies \citep[e.g.][]{2014A&A...570A...6S, 2016MNRAS.456L..35R, 2018MNRAS.478.2034R, 2019A&A...632A..59F,2019MNRAS.485.3794D,2020A&A...639L...9D}. This combination of codes allows for the simultaneous fit of the spectrum stellar features via a combination of the stellar models \citep[see][]{2004PASP..116..138C} as well as gaseous emission lines through independent Gaussian templates \citep[see][]{2006MNRAS.366.1151S, 2006MNRAS.369..529F}. In this analysis we make use of the MILES\footnote{The models are publicly available at \url{http://miles.iac.es} and
are based on the MILES empirical library \citep[][]{2006MNRAS.371..703S, 2011A&A...532A..95F}.} models constructed following the BaSTI \citep[][]{2004ApJ...612..168P} isochrones \citep[see][]{2015MNRAS.449.1177V, 2016MNRAS.463.3409V}. This approach allowed us to maximize the wavelength range covered by this analysis (including not only the continuum and stellar absorption features but also the regions affected by gaseous emission). 

\begin{figure}
\centering 
\includegraphics[width = 0.49\textwidth]{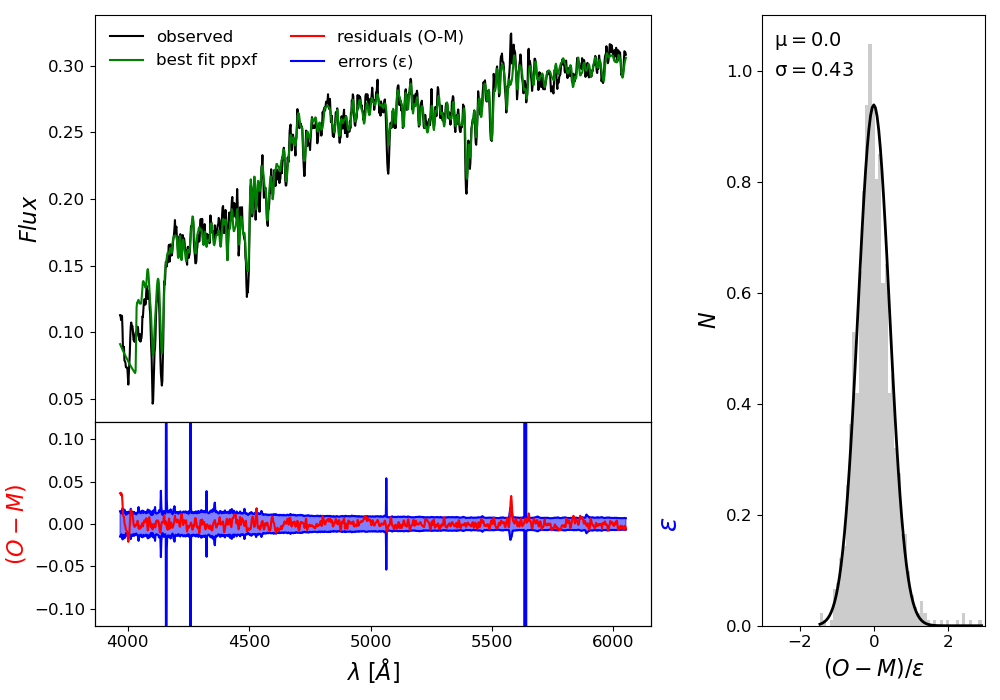} \\   
\caption{Example of full spectral fitting applied to the central Voronoi bin of CAVITY~40288. In the top-left panel, we overplot to the observed spectrum (black) and the best fit from {\tt pPXF} (green). The residuals of the fit (red) are compared to the propagated errors for the corresponding spectrum (blue) in the bottom-left panel. The right-hand panel shows a histogram of the ratio of the fitting residuals and the propagated errors including a Gaussian fit for this particular galaxy.}
\label{fig:err_prop_ind} 
\end{figure}

\begin{figure*}
\centering 
\includegraphics[width = 0.95\textwidth]{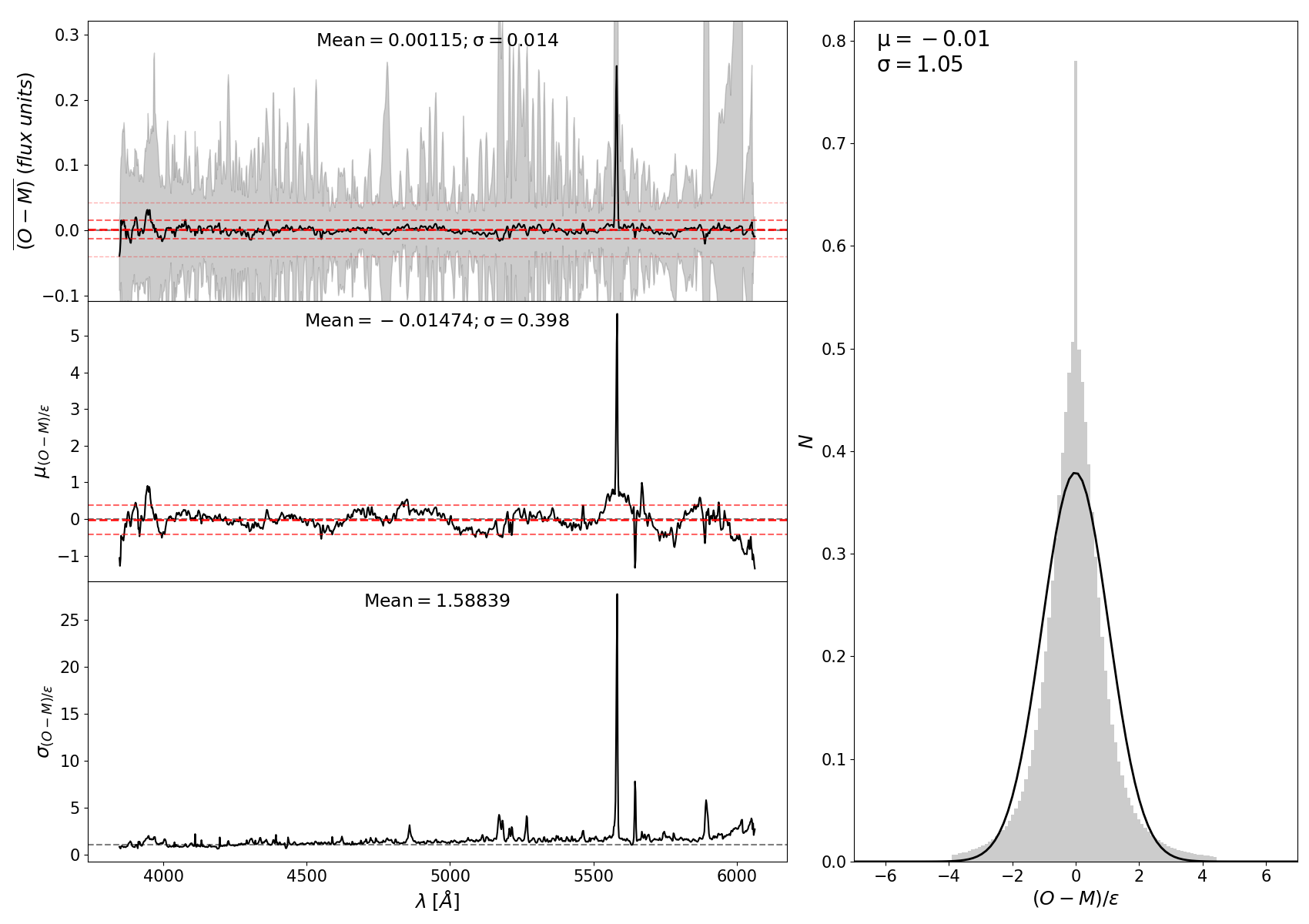} \\   
\caption{Comparison between the full spectrum fitting residuals and propagated errors in the CAVITY datacubes using a Voronoi spatial binning scheme. Top-left panel: Residual spectrum averaged over all Voronoid bins and all the CAVITY galaxies analysed. Middle-left panel: Mean of the ratio between the residuals and the propagated errors ($(O-M)/\epsilon$) as a function of wavelength. Bottom-left panel: Dispersion of $(O-M)/\epsilon$ as a function of wavelength. Right panel: Global distribution of the $(O-M)/\epsilon$. Red horizontal lines of gradual thickness are plotted in each panel for the mean value, mean $\pm$ $\sigma$, and mean $\pm$ 3$\sigma$. A Gaussian distribution representative of the mean and dispersion of the $(O-M)/\epsilon$ distribution is represented in the right-hand panel. In this analysis regions flagged as bad by the reduction pipeline have been masked from each individual spectrum as well as the 5577\AA~sky line.}
\label{fig:err_prop_vor} 
\end{figure*}

The spectra under analysis come from applying a Voronoi binning scheme \citep[][]{2003MNRAS.342..345C} to the CAVITY datacubes, requesting a target S/N of 30.  An example of this full spectral fitting approach applied to a typical spectrum from this spatial binning can be seen in Fig.~\ref{fig:err_prop_ind}. The distribution of the ratio between the residuals (O-M, Observed-Model, in red) and the propagated errors ($\epsilon$) is centred at zero, with a dispersion of $\sim$~0.4 (slightly overestimated in this case). Fig.~\ref{fig:err_prop_vor} summarises the global behaviour of all first 75 galaxies observed as part of the CAVITY project by stacking all analysed spectra. We confirm the validity of the data reduction error propagation, showing that the overall distribution of $(O-M)/\epsilon$ displays a mean value of -0.01 and a dispersion close to 1 (as expected). The run of the mean of the $(O-M)/\epsilon$ as a function of wavelength reflects that there are no clear systematic effects, although slightly larger deviations from the mean are found as we move to longer wavelengths (bottom-left panel). 

\section{CAVITY+: Project extension}
\label{cavityplus}

CAVITY was originally conceived as an IFS survey sampling galaxies inhabiting voids. However, with the goal of maximising the impact of the project, and mainly to fully address the proposed scientific goals, CAVITY has been expanded to include HI, CO and deep optical imaging of the IFS targets. 

In what follows, we explain the main generalities of the observations and reduction of each of the sub-projects. A complete description of all these complementary observational campaigns comprising CAVITY+ will be provided in future dedicated articles.

\subsection{Single dish CO}\label{sec:co}

A single dish CO spectroscopic survey was carried out on a sample of 106 CAVITY galaxies. This sample was selected from the galaxies with already available IFS data at the moment of submitting the proposal, with stellar masses above $10^9 \,\rm M_{\odot}$ (to avoid low-metallicity objects for which the Galactic CO-to-molecular gas mass conversion factor might not apply), and excluding quiescent systems (to ensure a moderately active star formation).

For the observations, we use the 30-m telescope of the Institut de Radioastronomie Millim\'etrique (IRAM) at Pico Veleta (Spain) to observe $^{12}$CO(1-0) and $^{12}$CO(2-1) line emission. The observations were carried out between December 2021 and February 2023. We used the E090 and the E230 EMIR bands simultaneously in combination with the autocorrelator FTS at a frequency resolution of 0.195 MHz (corresponding to a  velocity resolution of $\sim$ 0.5 \kms\ at CO(1--0) at the frequency of our observations) and with the autocorrelator WILMA with a frequency resolution of  2 MHz (corresponding to a  velocity resolution of  $\sim$ 5 \kms\  at CO(1--0)). The angular resolution was $\sim$ 22\arcsec\ and $\sim$ 11\arcsec\ for CO(1-0) and for CO(2-1), respectively. Each object was observed until it was detected with an S/N of at least five or until a root-mean-square noise (rms) of $\sim 1.5 $ mK (T$_{\rm mB}$) was achieved for the CO(1-0) line at a velocity resolution of 20 \kms . The on-source integration times per object ranged between 30~minutes and 4 hours. 

The data reduction was done with the CLASS program of the GILDAS software package\footnote{http://www.iram.fr/IRAMFR/GILDAS}, and involved the selection of good data, the subtraction of (linear) baselines from individual integration and the
averaging of the total spectra. Some observations taken with the FTS backend were affected by platforming, which was corrected using the {\it FtsPlatformingCorrection5.class} procedure provided by IRAM. 

After generating the final CO(1-0) and CO(2-1) emission spectra, we conducted a quality control analysis. This analysis involved inspecting each CO spectral profile and searching for possible errors and inconsistencies in the data reduction process. In particular, we checked the following points. (i) We visually inspected the spectra to make sure that the detections are convincing with no indications of artefacts. (ii) We confirmed that the baselines are flat, as wavy baselines can indicate an incorrect baseline subtraction or contamination with poor-quality data. (iii) We checked that the defined velocity window accurately constrained the entire CO emission and was centered on the recession velocity derived from the optical redshift. (iv) We confirmed the consistency between the CO(1-0) and CO(2-1) lines: the CO(1-0) line width is expected to be similar or larger than the CO(2-1) line width, which is due to the smaller beam at the higher frequency of CO(2-1) emitted from a smaller area.

To perform this analysis, we produced plots of the CO(1-0) and CO(2-1) lines at various velocity resolutions (10, 20, and 40 \kms ) to gain a different perspective of the data, which allowed us a more thorough inspection of the spectra. As mentioned before, one of the requirements was to observe each object until it was detected with an S/N of at least five or until a rms of $\sim1.5$ mK (T$_{\rm mB}$) was achieved for the CO(1-0) line at a velocity resolution of 20 \kms . A priori, this requirement ensured high quality for the CO data in general.  

Further details about the observations, sample selection, data reduction and the presentation of CO data will be published in Rodr{\'i}guez et al. (in prep).

The CO-CAVITY will be crucial to evaluate the gas reservoirs for future star formation in void galaxies. We will be able to compare the molecular gas mass to the SFR, the stellar mass, and the atomic mass (see Sec.~\ref{sec:hi}) in order to search for possible trends in the star-formation processes in different large-scale environments.

\subsection{Interferometric CO}

Interferometric CO(1-0) data have been secured from the Atacama Large Millimeter/submillimeter Array (ALMA) for a sub-sample of 41 galaxies at similar resolution than the IFS data (less than 1 kpc). The first observations happened in a Cycle 9 program and are ongoing with highest priority in a Cycle 10 program to complete the requested observations with over 40 hours. 

For the observations we use the 12-m array in three different configurations (C1, C3, and C4) corresponding to baselines from 0.16 to 0.78 km. We observe the low frequency CO(1-0) line (115 GHz) at a frequency resolution of 1.13 MHz (corresponding to a velocity resolution of $\sim$ 3 \kms ). 
The achieved angular resolutions of 1-1.5 \arcsec\ ensure linear resolutions better than 1 kpc ($\sim 350$ pc/arcsec at 4800 \kms\ and $\sim 900$ pc/arcsec at 10600 \kms ). The average S/N for a typical 1 arcsec$^2$ portion in any of the galaxies is 4.5 in the lower $25\%$ distribution, 10 in the lower $50\%$ distribution, and 20 in the lower $75\%$. The on-source integration time per object is around 30-50\,min. 

The data reduction is done with the ALMA pipeline (CASA, https://casa.nrao.edu/) and quality assurance following the standards of the observatory. Further details about the observations, sample selection, data reduction and the presentation of ALMA data will be published elsewhere.

This is the first interferometric CO survey of a statistically significant sample of void galaxies. This dataset, together with CAVITY IFS data, will allow us to elucidate the spatially resolved molecular gas properties and potential variations of star-formation efficacy in this largely unexplored type of galaxies.

\subsection{Atomic gas (HI)}\label{sec:hi}
Together with CO, HI data is key in order to have the full information of the neutral gas content necessary to understand the matter assembly and how the process of SF occurs in these galaxies. Thus, a dedicated observing campaign of 146 hours has been executed with the Green Bank Telescope (GBT) to observe 78 CAVITY galaxies with CO information from IRAM 30m (see Sec.~\ref{sec:co}) and for which no HI spectra was available from the literature (the ancillary data -- coming mostly from the ALFALFA project at Arecibo 305m -- will allow us to characterise the HI in the full CO-CAVITY sample). We used the L band with a bandwidth of 11.72 MHz. To observe the rest frequency of the HI line (1.420\,GHz), the (sky) topocentric frequencies for the different sources are between $1.378-1.355$\,GHz. We will obtain calibrated HI spectra with channel widths of about 20 \kms, and the HPBW of the telescope at these frequencies is $\sim 9\arcmin$. The on-source integration times per object range between 30 minutes to a few hours.

The data reduction is done with GBTIDL (https://gbtidl.nrao.edu/). Additional details about the observations, data reduction and the presentation of HI data will be published elsewhere.

These observations will allow us to characterise the HI mass of the galaxies. Furthermore, we will measure HI profile asymmetries and relate them to internal properties, such as the stellar component, bars, and morphological type, as well as external properties related to their galactic environment. 

\subsection{Deep optical imaging}

We are currently carrying out an extensive observational campaign with the 2.54-m {\it Isaac Newton} Telescope (INT) in order to provide deep imaging of the CAVITY galaxies. We use the Wide Field Camera (WFC), located at the primary focus, with a 4 CCD detector covering a FoV of approximately $34' \times 34'$ with a pixel scale of 0.333\arcsec. We use the SDSS $\rm g$ and $\rm r$ filters.

Observations have been allocated since semester 21A onward from both the NL-PC and CAT time allocation committees. To date, we have observed a total of 127 fields containing 141 CAVITY galaxies with IFS observations. In addition, a careful examination of the fields, to make the most out of each observing night, arose a total of 277 void galaxies (within the 15 voids that are targeted within the CAVITY project) with deep imaging in $\rm g$ and $\rm r$ bands reaching down to 30+ mag~arcsec$^{-2}$. All observations were carried out mainly in dark time, with some cases of grey time. Exposure times are nominally 1.5h per band, which can be increased if more exposure time is needed to reach the desired depths in case of moonlight.

The observational campaigns are designed with the aim of producing highly efficient flat-fielding at low surface brightness (LSB). For this purpose strong dithering patterns have been carried out that allow the science images themselves to be used to produce the flat image, a process that we detail below. The reduction is performed in what we call a `run'. This run consists of a set of images, typically the set of nights observed over a period of a few weeks, in which we assumed that the flat-fielding has no variation, so it can be corrected by a single flat image obtained from this run. For each night, the images are bias subtracted with a standard procedure. Subsequently, the bias subtracted images are heavily masked using \texttt{SExtractor} \citep{1996A&AS..117..393B} and \texttt{NoiseChisel} \citep{2015ApJS..220....1A}, normalised, and combined, producing a flat image. This flat is applied to the images producing the final reduced images.

\begin{figure}
\centering
    \includegraphics[width=1.0\columnwidth]{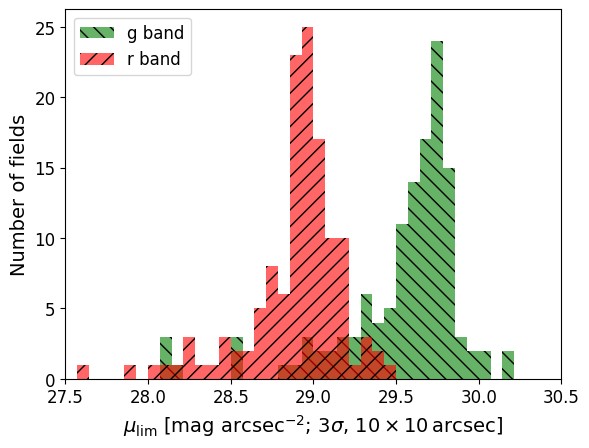}
    \caption{Histogram of SB limits in optical $\rm g$ and $\rm r$ bands measured at 3$\sigma$ in 10$\times$10 arcsec boxes for the INT data \citep[for details on the procedure see][]{2020A&A...644A..42R}. }
    \label{fig:Depth_cavity}
\end{figure}

For the combination of each of the observed fields, we make a provisional astrometric solution with the \texttt{astrometry.net} package \citep{2010AJ....139.1782L}, which is then refined with \texttt{SCAMP} \citep{2006ASPC..351..112B}. The co-adding process consists of a routine in which, through iterations of masking and sky subtraction, we converge to a final solution that we consider the final co-add. This methodology is considered the most robust for producing highly efficient images in the LSB regime \citep{2024MNRAS.528.4289W}. First we produce a seed co-add, which will be the beginning of the iteration. This co-add is produced by a constant sky subtraction for the whole image. This ensures that no over-subtraction of the data occurs in this seed co-add; however, there are certain gradients that are targeted to be removed by the process. To do this, after producing a mask in this co-add, this mask is applied to the individual exposures. We perform a sky subtraction on these individual exposures using Zernike polynomials \citep{1934MNRAS..94..377Z}. The motivation for using Zernike polynomials is to produce a sky subtraction compatible with the gradients of an optical system, minimising the potential oversubtraction of real sources present in the images \citep[see e.g., ][]{2023ApJ...948....4Z,2023A&A...679A.157R}. The images are then photometrically calibrated using SDSS and the noise of each individual image is calculated. The images are then combined by a 3$\sigma$ resistant mean; this mean is weighted by the individual noise of each image, producing the co-add that will be used as a new seed in this iterative process. This process of masking, sky subtraction and combination to obtain the co-add is repeated a number of times until a final co-add with excellent sky subtraction is achieved. We find that 3 iterations using low order Zernike polynomials, typically order 3, is sufficient. However, images with the presence of the moon or at low altitude require more iterations and a Zernike polynomial on order of four. Fig.~\ref{fig:Depth_cavity} presents histograms of SB limits in the $\rm r-$ and $\rm g-$bands. The depth analysis shows distributions in the g and r bands with peaks of approximately 29.7 mag~$\rm arcsec^{-2}$ in the $\rm g$ band and 29.0 mag~$\rm arcsec^{-2}$ in the $\rm r$ band, measured using a 3$\sigma$ in $10 \times 10$ arcsec boxes metric \citep[see][]{2020A&A...644A..42R}. We can notice tails in the distributions with shallower images, for instance below 29.5 mag~$\rm arcsec^{-2}$ in $\rm g-band$ and 28.7 mag~$\rm arcsec^{-2}$  in $\rm r-band$. These are exclusively due to the presence of the moon in the observations. As far as possible, future observations will add additional exposure times to these shallower images, in order to have homogeneity in the depths reached, avoiding images with shallower data. In the case of the computation of SB profiles, this translates to even fainter magnitudes (see Sect.~\ref{SB_prof_sect}).

To ensure the high quality of the deep imaging data, we carry out a visual inspection of the data when a set of new observations is processed by the deep imaging reduction pipeline. The revision is performed on each observed band per galaxy, with the redundancy of the pixels of each image. Because of the size of WFC/INT FoV, the full image covers an area much larger than the CAVITY galaxy size. This allowed us to look for any possible source of contamination by close galaxies, stars, or cirrus from the Milky Way. We therefore performed a visual inspection on the full images and on a zoomed-in area of the CAVITY galaxies. We defined a list of flags to identify any characteristic that could be related to an issue during observations, a problem derived from the reduction pipeline, or any characteristic that could be relevant when working with the data. We defined the following flags for the full images: i) STRIPES: if vertical or horizontal stripes, positive or negative ghost light, or artefacts are presented in the images; ii) INHOMOGENEOUS: if the images present irregular or not uniform background and fluctuations; iii) SPIKES: if there are stellar spikes, or bright spikes by other sources; iv) EXTERNAL\_OBJ: if there is any external object in the images (e.g., satellites, asteroids); v) REFLECTIONS: if there is any residual light affecting the full image; vi) DITHERING: if there is any issue with the dithering, like a pattern or lack of some pointings in the observations. When focusing on a CAVITY galaxy, we noted if the previous characteristics were affecting the galaxy as well, in addition to the following flags: i) STAR\_HALOS: if the galaxy image is affected by the stellar halo light from a close bright star; ii) STAR\_PUNCTUAL: if the galaxy image is affected by the light coming from a bright punctual star; iii) GALAXY\_HALOS: if the galaxy image is affected by the halo light from a close bright galaxy; iv) LSB\_FEATURES: if there are distinguishable LSB features such as interactions, streams, shells, or tails. %(Optional - An example of each flag is provided in Fig.)
In addition, we provide comments to report any extra characteristic or issue not covered by the flags (for instance serendipitous events).

After the first round of the deep imaging quality control, we found that around 30\% of the fields are affected by stripes inhomogeneities and spikes that are quantified and taken into account in the analysis of the data. These are the typical artefacts found in this type of dataset, and they do not impair the science analysis in most cases. In general, the field images are not much affected by external objects or reflections, and there is a problem related to dithering only in a few cases. When present, the reports were considered in each subsequent observation and reduction process. For the CAVITY galaxies, about 30\% of the images are affected by light coming from a bright nearby star, while 20\% are affected by a nearby bright galaxy. 

\begin{figure*}
\centering 
\includegraphics[width = 2.0\columnwidth]{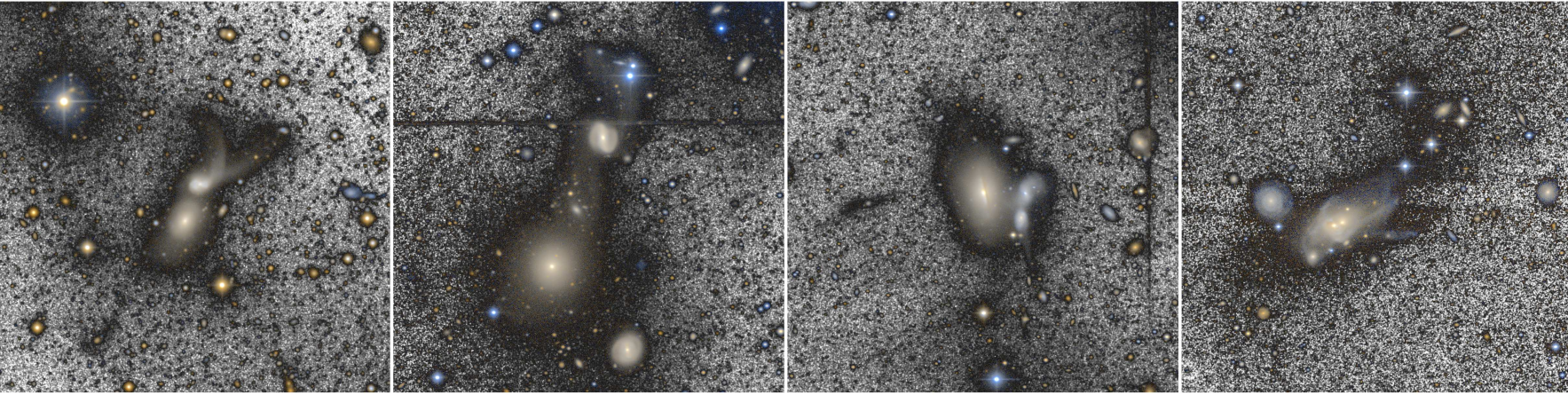} \\   
%\caption{Composed image for CAVITY 33694 and CAVITY 33695 galaxies, an interactive system with an undetectable tidal tail in shallower data, but detected in the INT deep images, with $\rm \mu_g\,\sim\, 28\, mag\, arcsec^{-2}$ in the g-band. The image in composed of the HSC? three-colour image and INT g-band.}
\caption{Composed image for a sample of CAVITY galaxies with undetectable tidal tails in shallower data but detected in the CAVITY INT deep images, with $\rm \mu_g\,\sim\, 28\, mag\, arcsec^{-2}$ in the $\rm g$-band. The colour image was constructed with the $\rm g$ and $\rm r$ bands. The dark background with the sum image $\rm g+r$ in high contrast.}
\label{fig:int_deep_lsb} 
\end{figure*}

The deep-imaging CAVITY follow-up will allow us to detect LSB structures indicating possible interactions, low-mass companions, or other local environment indicators that can help characterise their influence in the evolution of void galaxies. In the first round of observations, we detected LSB features in 26\% of the CAVITY galaxies. An example of LSB features found in a few CAVITY galaxies is shown in Fig.~\ref{fig:int_deep_lsb}. For a more in-depth assessment of the science to be done with these data, in Sect.~\ref{SB_prof_sect} we present some examples of SB profile determination.

\section{CAVITY science}
\label{science}
The IFS database for the 300 galaxies of the CAVITY survey, together with all additional information coming from the CAVITY+ extended project, will offer a unique opportunity to explore the properties of galaxies residing in cosmic voids and will allow the community to address a large number of astrophysical questions. 
Within the expertise of the CAVITY team, there is an interest to explore the following aspects of galaxies in voids: the baryonic mass assembly; the impact of the large-scale environment on the gas accretion; the sSFR and the molecular and atomic gas content of void galaxies; the merger and accretion histories of galaxies from the light distribution in the outer parts of galaxies; the influence of the local versus the large-scale environment on general galaxy properties; the effects of the large-scale structure in the prevalence of AGNs, their properties, and their role in quenching or enhancing star formation in void galaxies; and the properties and formation of dwarf galaxies in voids.

We have already carried out some preparatory analysis of the SDSS spectroscopic data of the central parts of the galaxies within the CAVITY parent sample \citep[][]{2023arXiv231011412D, 2023Natur.619..269D} where we concluded that the stellar mass assembly of void galaxies occurs slower than in filaments and walls, and much slower than in clusters, pointing to different physical drivers of the galaxy evolution for the different large-scale environments. Furthermore, the analysis of the stellar mass-metallicity relation on the same sample shows that the enrichment also occurs differently in the different large-scale environments. These works, although ground-breaking, lack the 2D information crucial to discern among the different physical processes that drive the different SFHs at different densities. 
 
In this section, as an example of the power of combining CAVITY and CAVITY+ data and to present the type of products that will be produced by the collaboration, we have selected a sample of four CAVITY galaxies that will serve as pilot sample for the CAVITY science. These galaxies are CAVITY10668, CAVITY48125, CAVITY49137, and CAVITY59902 (see Table~\ref{table:1}). For all galaxies we have IFS, INT deep imaging, and single-dish CO data, except for CAVITY49137, a barred galaxy in a triple system with only IFS data available so far. The selected galaxies show similar absolute magnitudes in $\rm r$ band (ranging from -19.53 to -20.37 mag) and a wide range of $\rm g-r$ colours (from 0.549 to 0.852 mag). The purpose here is to show the power of the CAVITY data to gather kinematic, chemical, and physical properties of their gas and stars to be linked to local interactions, global environmental effects, and secular evolution. The characterisation of these properties on the full CAVITY sample will enable the main science drivers of the project to be addressed. For this purpose, we analyse and present here the data products from the IFS, the deep imaging, and the CO survey for these four galaxies. 
 
\subsection{Gas and stellar population properties}

To analyse the CAVITY datacubes presented in this work, we have used {\tt Pipe3D} \citep{2016RMxAA..52...21S, 2016RMxAA..52..171S}, which was developed to characterise the stellar and ionised gas properties as derived from optical spectra, in particular, it has been optimised to work with optical IFS data. Specifically, we have made use of {\tt pyPipe3D}, the new publicly available implementation for {\tt Python} \citep{2022NewA...9701895L}.

In order to derive 2D maps of the stellar properties, {\tt pyPipe3D} first performs a spatial binning called Continuum Segmentation binning (CS-binning) to increase the S/N of the spectra and obtain an accurate estimation of the stellar contribution. This binning scheme combines a continuity criterion for the SB (applying a specified maximum difference in the flux intensity to aggregate adjacent spaxels) and a goal for the S/N ratio (in our case, 50). Then, the code fits each spectrum by a linear combination of single stellar population synthesis model (SSP) spectra after correcting for the appropriate systemic velocity and velocity dispersion and taking into account the effects of dust attenuation \citep[][with $R_V=3.1$]{1989ApJ...345..245C}. As SSP templates we use the MILES \citep[][]{2015MNRAS.449.1177V} stellar synthesis models (BASE models) using the BaSTI isochrones \citep[][]{2004ApJ...612..168P}. We use the Kroupa Universal Initial Mass Function \citep[IMF,][]{2001MNRAS.322..231K}, with an IMF slope of 1.3. 

\begin{figure*}
\centering 
\includegraphics[width = \textwidth]{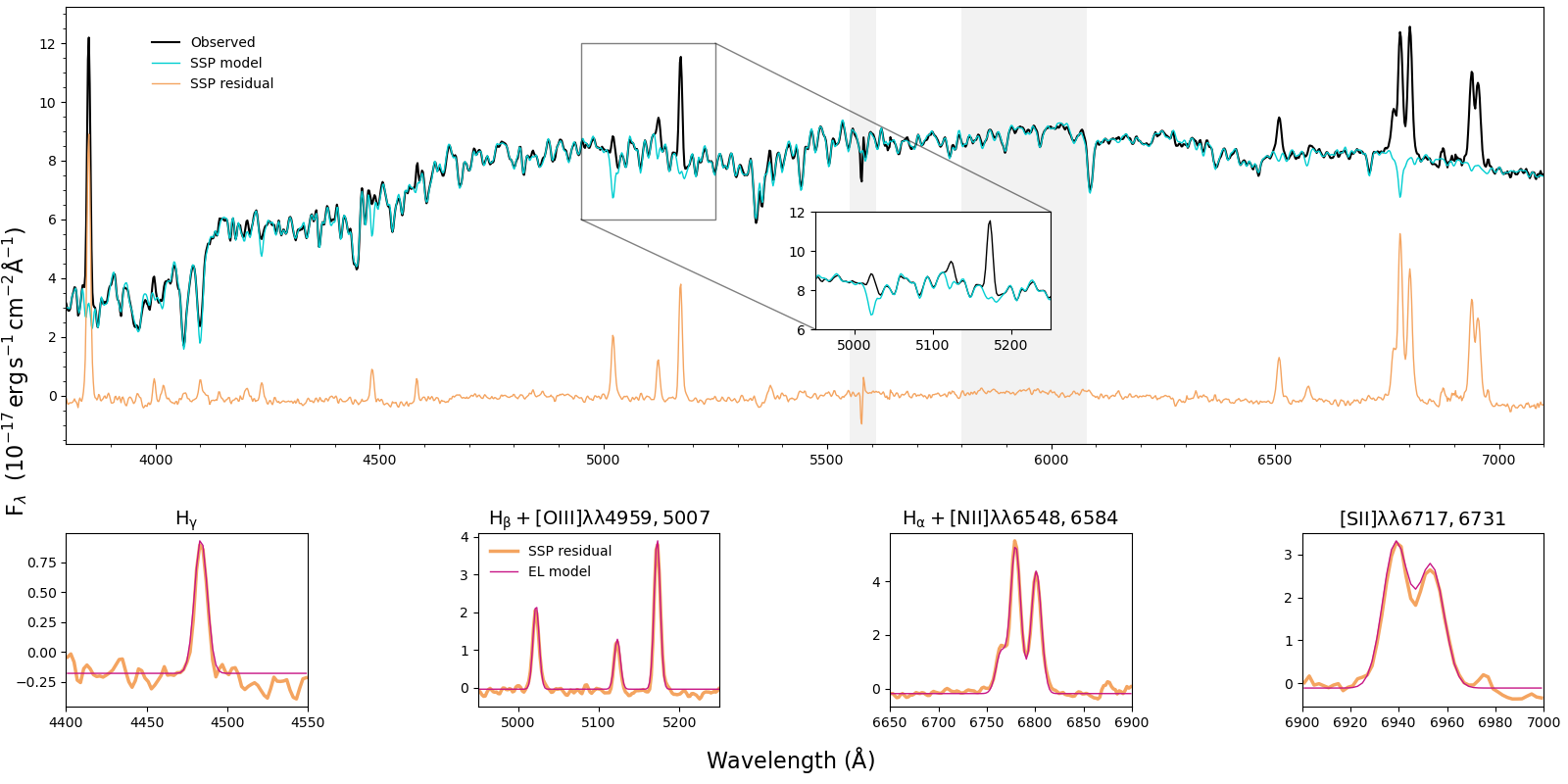} \\   
\caption{Example of a {\tt pyPipe3D} fit to the central spectrum ($5\arcsec$ diameter) of CAVITY49137 extracted from the IFS datacube. In the top panel, the black line represents the original spectrum, along with the best fitted stellar population (SSP model, cyan). The pure emission line spectrum after subtracting the best stellar population model is shown as a solid yellow line (SSP residual). The inset is focused on the H$\beta$-[OIII] spectral region to highlight the quality of the fit. The vertical shaded regions indicate the wavelength ranges masked during the fitting process. The bottom panels show some examples of emission line fittings (emission line (EL) model, red line).}
\label{fig:pipe3d} 
\end{figure*}

Once we have an estimation of the stellar continuum, {\tt pyPipe3D} performs a procedure called {\em dezonification} to decouple the analysis of the emission lines from the spatial binning required for the stellar populations, taking into account the relative contribution of each spaxel to the spatial bin in which it is aggregated \citep{2013A&A...557A..86C}. The {\em dezonification} map is multiplied to the stellar model of the datacube and the resulting cube is spatially smoothed with a Gaussian kernel to obtain a continuous model of the underlying stellar population. Finally, this model is subtracted
from the original datacube providing a pure emission line cube. For the analysis of the emission lines, the algorithm performs a multi-component fitting on the pure gas cubes using a single Gaussian function per emission line plus a low-order polynomial function. The treatment of the errors is done via a Monte Carlo simulation, and it includes the propagation of the uncertainties from the subtraction of the underlying stellar population to the parameters derived for the emission lines. A full description of the whole procedure performed by {\tt Pipe3D} can be found in \citet{2022NewA...9701895L} and \citet{2016RMxAA..52..171S}. Figure~\ref{fig:pipe3d} illustrates the quality of the performance of {\tt pyPipe3D} showing as an example the fit of the central spectrum ($5\arcsec$ diameter) of CAVITY49137. 

 %The emission line maps are derived subtracting the stellar continuum obtained from a combination of single stellar populations (SSP)  with different ages and metallicities after correcting for the appropriate systemic velocity and velocity dispersion and taking into account the effects of dust attenuation \citet{1989ApJ...345..245C}. Once the stellar component is subtracted, Pipe3D proceeds to measure the emission lines performing a multi-component fitting using a single Gaussian function per emission line plus a low order polynomial function. The emission maps presented here correspond to star formation regions selected using the BPT diagram \citet{1981PASP...93....5B}.

\begin{table*}
\caption{Properties of void galaxies.} % title of Table
\label{table:1}      % is used to refer this table in the text
\centering                          % used for centering table
\begin{tabular}{c c c c c}        % centered columns (4 columns)
\hline\hline                 % inserts double horizontal lines
CAVITY ID & redshift & $\rm g-r$  & $\rm r$-band & Stellar mass\\    % table heading 
 & & absolute magnitude& & \\
& & [mag]&[mag]& [log(M$_{\odot}$)]\\
\hline                        % inserts single horizontal line
   10668 & 0.03773 & 0.640 & -20.366&9.840\\      % inserting body of the table
   48125 & 0.03293 & 0.714 & -20.021&10.399\\
   49137 & 0.03308 & 0.852 & -20.005&10.486\\
   59902 & 0.02993 & 0.549 & -19.529& 9.909\\
\hline                                   %inserts single line
\end{tabular}
\end{table*}

\begin{table*}
\caption{Average stellar and
gaseous properties of void galaxies.}             % title of Table
\label{table:2}      % is used to refer this table in the text
\centering                          % used for centering table
\begin{tabular}{cccccc}        % centered columns (4 columns)
\hline\hline                 % inserts double horizontal lines
CAVITY ID & Molecular mass& $\langle$ log(age[yr])$\rangle$ & {[Z/H]}$_{\rm R_{eff}}$  & 12+log([O/H])$_{\rm R_{eff}}$ & $\sigma_{\rm R_{eff}}$ \\   
& log(M$_{\odot})$& & & &[kms$^{-1}$]\\
% table heading 
\hline                        % inserts single horizontal line
   10668 &$9.50 \pm 0.04$  & $9.0 \pm 0.2$ & $-0.3 \pm 0.3$ & $8.56 \pm 0.03$ & $150 \pm 20$ \\      % inserting body of the table
   48125 & $9.32 \pm 0.04$ & $9.1 \pm 0.2$ & $-0.2 \pm 0.2$ & $8.57 \pm 0.05$ & $160 \pm 20$ \\
   49137 &--  & $9.6 \pm 0.2$ & $-0.1 \pm 0.2$ & $8.47 \pm 0.04$ & $50 \pm 20$ \\
   59902 &  $8.85 \pm 0.06$ & $8.8 \pm 0.2$ & $-0.3 \pm 0.2$ & $8.50 \pm 0.04$ & $120 \pm 10$ \\
\hline                                   %inserts single line
\end{tabular}
\end{table*}

Figure~\ref{fig:results_ifu} shows typical data products of the IFS cubes for the selected sample of four void galaxies applying the above explained methodology. The ionised gas abundance maps have been derived from the {\tt pyPipe3D} emission line intensity maps using the O3N2 strong-line abundance indicator with the \cite{2013A&A...559A.114M} calibrator. {The light-weighted [Z/H] maps have been directly obtained from {\tt pyPipe3D} as described in \cite[][see Eq. 12]{2022NewA...9701895L}.} For the objects that also have deep imaging and CO data available, the CO(2-1) and CO(1-0) spectroscopic profiles as well as the optical deep images and SB profiles are shown in Figs.~\ref{fig:results_co} and~\ref{fig:SB_profiles}.

Figure~\ref{fig:results_s2n} shows a quick quality assessment of the IFS datacubes included here, we present S/N maps (and histograms) of the galaxies with contours indicating an S/N level of 10 and 30, respectively. According to experience, this would correspond to the area within the galaxy for which we can provide reliable kinematics and stellar parameters. The quality of the CAVITY datacubes will enable the gathering of spectroscopic information, reaching an SB of around 23~mag~arcsec$^{-2}$ with an S/N $\sim$~30 and down to nearly 25~mag~arcsec$^{-2}$ with an S/N $\sim$~10. 

Table~\ref{table:2} shows some average properties derived from the previous analysis. Among other quantities, the average stellar ages and metallicities at one effective radius (R$_{\rm eff}$), as well as the ionised gas averaged metallicity at the same radius are shown. The average stellar metallicities and ages are in agreement with the MZ$_{\star}$R (stellar mass-metallicity relation) for void galaxies with long-time SFHs, defined as those galaxies that formed a low fraction of their stellar mass (< 21.4\%) 12.5 Gyr ago, but formed stars uniformly over time \citep{2023arXiv231011412D}.

According to the criteria of local environment defined in \citet{2015A&A...578A.110A} two of the galaxies are in isolation, CAVITY48125, and CAVITY59902; whereas CAVITY10668 is part of an isolated pair and CAVITY49137 lives in a widely separated triple system. The two isolated systems, as expected, show symmetric kinematic, stellar age, and metallicity maps. In particular, we observed a clear alignment between gas and stellar kinematics, an aspect that we do not see in CAVITY10668 (especially) or CAVITY49137, where some differences can be seen between the stellar and gaseous kinematics. INT imaging of these two isolated systems display quite symmetric discs up to the outermost regions, though some subtle asymmetries are detected by our deep imaging in their outskirts. 

CAVITY10668 is an interacting pair and the velocity maps show a clear misalignment between the ionised gas and the stellar distribution, as well as asymmetries in the age and metallicity distributions. The ionised-gas distribution is aligned with the elongated light distribution of the galaxy itself, whereas the stellar kinematics is more oriented towards its companion\footnote{At this point we must say that CAVITY10669, the companion to CAVITY10668 and seen in the INT deep imaging, was also observed as part of CAVITY but with poor weather conditions. We plan to reobserve this system in order to fully study this interesting pair.}. The ionised-gas metallicity map shows an increase in metallicity towards the tail with the interacting galaxy (to the north of the image). The stellar metallicity shows a lowering of the metallicity towards the central regions and an increment towards the north-west of the image, towards the location where the companion is placed. The CO spectra for this galaxy is centrally concentrated at zero velocity, possibly linked to a concentration of molecular gas in the central regions. This interpretation would be compatible with the ionised-gas kinematic 2D map (see Figure~\ref{fig:results_ifu}) that shows zero velocity contours only in the central regions. The kinematic misalignment, the metallicity asymmetries, and the CO concentration can all be consequence of the interaction state in which this galaxy is found.  
The other non-isolated system, CAVITY49137, shows also asymmetries in all the maps, no deep imaging is available for this galaxy, but a visual inspection of the DECALS image \citep{2019AJ....157..168D} shows some signs of lopsidedness in the light distribution of this barred galaxy that needs to be confirmed with deeper imaging. The twisted iso-velocity contours present in the central parts of the gas kinematics are probably due to the presence of a clear bar \citep[e.g.][]{2015MNRAS.451..936S}. The CO(2-1) and CO(1-0) data are still not available for this galaxy. 

The few examples presented here are proof of the capability of the CAVITY survey to characterise the spatial stellar and gas properties of galaxies inhabiting voids for the first time in a statistical sample, a first study on the spatially resolved stellar populations of 118 CAVITY galaxies has been recently accepted for publication \citep{2024arXiv240410823C}. The combination of deep optical imaging and CO observations will allow us to determine the effect of the local and global environment on the galaxy properties, providing a unique view of galaxies in cosmic voids.
 
%\begin{figure*}
%\centering 
%\includegraphics[width = \textwidth]{int_plot_no_prof.png} \\   
%\caption{Deep imaging for the CAVITY galaxies. From left to right g-band images from our INT deep images for CAVITY10668, CAVITY48125, and CAVITY59902. We overlay the same contours as in Fig.~\ref{fig:results_ifu} from the analysis of the IFU data.}
%\label{fig:results_int} 
%\end{figure*}

%\begin{figure*}
%\centering 
%\includegraphics[width = \textwidth]{int_plot_decals.png} \\   
%\caption{Deep imaging for the CAVITY galaxies. Top panels: g-band images from our INT deep images. We overlay the same contours as in Fig.~\ref{fig:results_ifu} from the analysis of the IFU data. Bottom panels: g-band surface brightness profiles as computed from elliptical fitting comparing our INT observations (black) with DECALS photometry (red).}
%\label{fig:results_int} 
%\end{figure*}

\subsection{Deep imaging and light distributions}
\label{SB_prof_sect}

A visual inspection of the deep images we are gathering already reveals the variety of galaxies in cosmic voids (see Fig.~\ref{fig:int_deep_lsb}). These deep images from the CAVITY+ INT campaign are clear evidence that void galaxies, contrary to the intuition, are not necessarily isolated systems. Indeed, the four galaxies selected for this presentation paper inhabit different local environments as well, highlighting this fact. Apart from allowing us to study the more local environment of the CAVITY galaxies, the deep imaging data unveiled a large number of LSB structures,\footnote{All of these features and structures will be thoroughly analysed in a future work within the CAVITY collaboration.} providing a perfect complement to the IFS data and, in some cases, CO information. 

In principle, one might expect that internal evolutionary effects are more important in void galaxies than environmental effects as they inhabit these low-density regions of the Universe. However, the effects of minor mergers should not be neglected. As such, a thorough characterisation of the outer parts, including the study of the SB profiles and asymmetries, can shed light into the past accretion histories of the analysed galaxies. The large number of galaxies that are being observed as part of this deep imaging campaign enables an unprecedented comparison between accretion histories of galaxies in different environments. This, in combination with the IFS and CO data (and other ancillary comprising CAVITY+; see Sec.\ref{cavityplus}) will complete the puzzle of how these galaxies have formed and evolved.

In line with the previous section, here we analyse the SB profiles of CAVITY10668, CAVITY48125, and CAVITY59902 (the ones with INT data so far; see Fig.~\ref{fig:SB_profiles}). In the case of galaxies in interaction, such as CAVITY10668, low SB features and asymmetries abound, hindering the interpretation of their SB profiles. In the case of more isolated galaxies (e.g. CAVITY48125 and CAVITY59902), analysis of their outer regions is enabled. Figure~\ref{fig:SB_profiles} exemplifies our process of computing SB profiles and compare the profiles from the INT campaign to those from DECALS. 

In short, we measured the SB profiles of each of the galaxies in the \textit{g}-band image. Proper local background subtraction and image masking need to be done to study the low SB regions. We construct masks for each image using the software \texttt{SExtractor} \citep[][v.2.25.0]{1996A&AS..117..393B}. We combine two different masks, the first one with the parameters optimised for detecting extended sources (larger background size and intermediate threshold) and the other one optimised for point source detection (lower background size and threshold). Additionally, we mask all peaks detected in the residual mask of the image minus a Gaussian smoothed version of the image that are larger than the PSF (FWHM $\sim 1.2$ arcsec). Finally, we visually inspect the mask and refine any region by hand.
With all the contaminant sources masked properly, we measure the local background in boxes around the galaxy. We set the distance of the boxes to be larger than the region where an elliptical profile of the galaxy is flattened. Then, we typically placed around $\sim 10 - 20$ boxes of at least $50$ not-masked pixels around the galaxy and we measure the mean of all pixels not masked inside the boxes rejecting pixels above $2.5$ times the standard deviation. Then, we set the local value to be the mean of all boxes. The dominant uncertainty at this level is the value of the local background, so we measure the error on this value as the standard deviation of all the boxes. We construct SB radial profiles using the implementation in \texttt{astropy} \citep[][v5.1]{astropy, astropy2018} of the method described by \citet{Jedrzejewski1987}. This method fits elliptical apertures at each radial bin. We increase the spacing between each aperture logarithmically, using $2\%$ bigger radii than the previous aperture.

As a general remark, especially in the cases of CAVITY10668 and CAVITY59902, we can see how we reach around 1 to 2 ~mag~arcsec$^{-2}$ deeper than with the same approach applied to DECALS data. For CAVITY48125, some of the exposures were affected by residual moonlight restricting the depth that we reach. Even in this case, the computed profile perfectly traces the one obtained using DECALS data, matching its depth. For the cases of the isolated systems (CAVITY48125 and CAVITY59902) we can see how reliably reaching a magnitude of $\sim$~30~mag~arcsec$^{-2}$ is a reality, starting to grasp emission from the halo of the analysed galaxies. In the case of CAVITY10668, the outermost regions are clearly affected by the presence of the companion galaxy. Interestingly, in the case of CAVITY48125 and CAVITY59902 we observed clear downbending profiles \citep[][]{2006A&A...454..759P}. However, the origin of both profiles seem to be different. Whereas in the case of CAVITY48125 the so-called break is located at the end of the spiral pattern, and probably linked to it \citep[][]{2006ApJ...645..209D}, for CAVITY59902 it does not seem to be clearly related to any morphological feature, and thus, probably related to its mass distribution (due to a change in the angular momentum of the disc or a star-formation threshold; see, e.g., \citealt[][]{2008ApJ...675L..65R, 2009MNRAS.398..591S}). These profiles evidence the importance of relying on deep imaging data in order to properly characterise the outskirts of the galaxies, more susceptible to environmental effects \citep{2023A&A...677A.117S}.

\section{Conclusions}
The CAVITY survey aims to understand the influence of the large-scale environment on galaxy formation and evolution. For that, CAVITY is gathering new observations that are crucial to unveil the history of the observed galaxies. To date, CAVITY is the largest and most complete survey on the detailed properties of void galaxies. The survey was initiated as an IFS survey using the PMAS/PPAK spectrograph at the 3.5 meter telescope at the Observatory of Calar Alto, Spain. This initial survey has been extended to a CO, HI, and optical deep imaging follow-up of the sample. The single dish CO(1-0) and CO(2-1) emission observations are from the IRAM 30m telescope, the resolved CO(1-0) data are collected with ALMA, the HI spectra are obtained with the Green Bank Telescope, and the deep $\rm g$- and $\rm r$-band imaging is being carried out at the Wide-Field-Camera at the {\it Isaac Newton} Telescope at La Palma Observatory. These complementary datasets comprise the CAVITY+ project.

In this work, we have presented the parent sample from which the 300 galaxies that will form CAVITY have been selected. These 300 galaxies populate 15 nearby voids, sampling across the voids with around 20 galaxies per void and populating the colour-magnitude diagram. These criteria ensures that CAVITY will offer a complete view of the properties of galaxies in voids. A quick outline of the observations, data reductions, and quality of the data has been given in this paper, although separate in-depth papers are planned in order to fully present the CAVITY and CAVITY+ data.

To give the community a feeling of the quality of the data and products that will be released in the CAVITY project, apart from the assessment carried out in Sect.~\ref{QC}, we have presented data-products for four galaxies. We have shown that we will provide 2D stellar ages and metallicities as well as ionised gas properties and stellar and gas kinematics up to the outer parts of these galaxies. We have presented results for two isolated galaxies, an interacting one and an example of a small group residing in cosmic voids (proving that some galaxies in voids present a rich local environment). These results show that the kinematic, stellar, and gas property substructures, asymmetries, and gradients can be derived from the data. These spatially resolved parameters will be key to constraining the evolution of galaxies residing in cosmic voids. Deep imaging provides extra information about the possible local environment processes driving some internal properties (beyond the FoV of the IFS data) as well as SB profiles down to 30 mag/arcsec$^2$. The addition of CO and HI data will help in evaluating the gas reservoirs and the star-formation efficiency in void galaxies. 
A first data release will be open to the community in July 2024, offering datacubes for a sample of around 100 galaxies through an easy-to-access database platform based on the Daiquiri framework at the survey's web page.\footnote{\url{http://cavity.caha.es}}

\begin{acknowledgements}
Based on observations collected at the Centro Astron\'omico Hispano en Andaluc\'ia (CAHA) at Calar Alto, operated jointly by Junta de Andaluc\'ia and Consejo Superior de Investigaciones Cient\'ificas (IAA-CSIC). We acknowledge financial support by the research projects AYA2017-84897-P, PID2020-113689GB-I00, and PID2020-114414GB-I00, financed by MCIN/AEI/10.13039/501100011033, the project A-FQM-510-UGR20 financed from FEDER/Junta de Andaluc\'ia-Consejer\'ia de Transformaci\'on Econ\'omica, Industria, Conocimiento y Universidades/Proyecto and by the grants P20-00334 and FQM108, financed by the Junta de Andaluc\'ia (Spain). TRL acknowledges support from Juan de la Cierva fellowship (IJC2020-043742-I). LSM acknowledges support from Juan de la Cierva fellowship (IJC2019-041527-I).
DE acknowledges support from a Beatriz Galindo senior fellowship (BG20/00224) from the Spanish Ministry of Science and Innovation.
RGD, RGB, and AC acknowledge financial support from the State Agency for Research of the Spanish MCIU through ‘Center of Excellence Severo Ochoa’ award to the Instituto de Astrof\'isica de Andaluc\'ia, CEX2021-001131-S, funded by MCIN/AEI/10.13039/501100011033, and to financial support from the projects PID-2019-109067-GB100 and PID2022-141755NB-I00. HC also acknowledges support from the Institut Universitaire de France and from the CNES. IMC acknowledges support from ANID programme FONDECYT Postdoctorado 3230653.
JF-B acknowledges support from the PID2022-140869NB-I00 grant from the Spanish Ministry of Science and Innovation. VQ,SP, and MH acknowledge that this work has been supported by the Agencia Estatal de Investigación Española (AEI; grant PID2022-138855NB-C33), by the Ministerio de Ciencia e Innovaci\'on (MCIN) within the Plan de Recuperaci\'on, Transformaci\'on y Resiliencia del Gobierno de España through the project ASFAE/2022/001, with funding from European Union NextGenerationEU (PRTR-C17.I1), and by the Generalitat Valenciana (grant PROMETEO CIPROM/2022/49).
PSB acknowledges financial support from the from the Spanish Ministry of Science under the projects with references:  PID2019-107427GB-C31 and PID2022-138855NB-C31. PVG acknowledges that the project that gave rise to these results received the support of a fellowship from “la Caixa” Foundation (ID 100010434). The fellowship code is B005800. AFM acknowledges support from RYC2021-031099-I and PID2021-123313NA-I00 of MICIN/AEI/10.13039/501100011033/FEDER, UE, NextGenerationEU/PRT. 
M.A-F. acknowledges support from the Emergia program (EMERGIA20-38888) from Consejer\'ia de Universidad, Investigaci\'on e Innovaci\'on de la Junta de Andaluc\'ia. JR acknowledges funding from University of La Laguna through the Margarita Salas Program from the Spanish Ministry of Universities ref. UNI/551/2021-May 26, and under the EU Next Generation.  BB acknowledges financial from the Grant AST22$\_$4.4, funded by Consejer\'ia de Universidad, Investigaci\'on e Innovaci\'on and Gobierno de Espa\~na and Uni\'on Europea – NextGenerationEU, and by the research project PID2020-113689GB-I00 financed by MCIN/AEI/10.13039/501100011033. GTR acknowledges financial support from the research project PRE2021-098736, funded by MCIN/AEI/10.13039/501100011033 and FSE$+$.

This research made use of
astropy, a community-developed core python (http://www.python.org)
package for Astronomy (Astropy Collaboration 2013); ipython (P\'erez \&
Granger 2007); matplotlib (Hunter 2007); numpy (Walt et al. 2011); scipy
(Jones et al. 2001); and topcat (Taylor 2005).

This research has made use of the
NASA/IPAC Extragalactic Database, operated by the Jet Propulsion Laboratory
of the California Institute of Technology, un centract with the National Aero-
nautics and Space Administration. Funding for SDSS-III has been provided by
the Alfred P. Sloan Foundation, the Participating Institutions, the National Sci-
ence Foundation, and the U.S. Department of Energy Office of Science. The
SDSS-III Web site is http://www.sdss3.org/. The SDSS-IV site is http:
//www.sdss.org.

This publication was based on observations collected under the CAVITY legacy project.
\end{acknowledgements}

\bibliographystyle{aa} % style aa.bst
\bibliography{references.bib}

\begin{thebibliography}{104}
\expandafter\ifx\csname natexlab\endcsname\relax\def\natexlab#1{#1}\fi

\bibitem[{{Abazajian} {et~al.}(2009){Abazajian}, {Adelman-McCarthy}, \&
  {Ag{\"u}eros}}]{2009ApJS..182..543A}
{Abazajian}, K.~N., {Adelman-McCarthy}, J.~K., \& {Ag{\"u}eros}. 2009, \apjs,
  182, 543

\bibitem[{{Ahumada} {et~al.}(2020){Ahumada}, {Allende Prieto}, {Almeida},
  {Anders}, {Anderson}, {Andrews}, {Anguiano}, {Arcodia}, {Armengaud},
  {Aubert}, {Avila}, {Avila-Reese}, {Badenes}, {Balland}, {Barger},
  {Barrera-Ballesteros}, {Basu}, {Bautista}, {Beaton}, {Beers}, {Benavides},
  {Bender}, {Bernardi}, {Bershady}, {Beutler}, {Bidin}, {Bird}, {Bizyaev},
  {Blanc}, {Blanton}, {Boquien}, {Borissova}, {Bovy}, {Brandt}, {Brinkmann},
  {Brownstein}, {Bundy}, {Bureau}, {Burgasser}, {Burtin}, {Cano-D{\'\i}az},
  {Capasso}, {Cappellari}, {Carrera}, {Chabanier}, {Chaplin}, {Chapman},
  {Cherinka}, {Chiappini}, {Doohyun Choi}, {Chojnowski}, {Chung}, {Clerc},
  {Coffey}, {Comerford}, {Comparat}, {da Costa}, {Cousinou}, {Covey}, {Crane},
  {Cunha}, {Ilha}, {Dai}, {Damsted}, {Darling}, {Davidson}, {Davies}, {Dawson},
  {De}, {de la Macorra}, {De Lee}, {Queiroz}, {Deconto Machado}, {de la Torre},
  {Dell'Agli}, {du Mas des Bourboux}, {Diamond-Stanic}, {Dillon}, {Donor},
  {Drory}, {Duckworth}, {Dwelly}, {Ebelke}, {Eftekharzadeh}, {Davis Eigenbrot},
  {Elsworth}, {Eracleous}, {Erfanianfar}, {Escoffier}, {Fan}, {Farr},
  {Fern{\'a}ndez-Trincado}, {Feuillet}, {Finoguenov}, {Fofie},
  {Fraser-McKelvie}, {Frinchaboy}, {Fromenteau}, {Fu}, {Galbany}, {Garcia},
  {Garc{\'\i}a-Hern{\'a}ndez}, {Garma Oehmichen}, {Ge}, {Geimba Maia},
  {Geisler}, {Gelfand}, {Goddy}, {Gonzalez-Perez}, {Grabowski}, {Green},
  {Grier}, {Guo}, {Guy}, {Harding}, {Hasselquist}, {Hawken}, {Hayes}, {Hearty},
  {Hekker}, {Hogg}, {Holtzman}, {Horta}, {Hou}, {Hsieh}, {Huber}, {Hunt}, {Ider
  Chitham}, {Imig}, {Jaber}, {Jimenez Angel}, {Johnson}, {Jones},
  {J{\"o}nsson}, {Jullo}, {Kim}, {Kinemuchi}, {Kirkpatrick}, {Kite}, {Klaene},
  {Kneib}, {Kollmeier}, {Kong}, {Kounkel}, {Krishnarao}, {Lacerna}, {Lan},
  {Lane}, {Law}, {Le Goff}, {Leung}, {Lewis}, {Li}, {Lian}, {Lin}, {Long},
  {Longa-Pe{\~n}a}, {Lundgren}, {Lyke}, {Mackereth}, {MacLeod}, {Majewski},
  {Manchado}, {Maraston}, {Martini}, {Masseron}, {Masters}, {Mathur},
  {McDermid}, {Merloni}, {Merrifield}, {M{\'e}sz{\'a}ros}, {Miglio}, {Minniti},
  {Minsley}, {Miyaji}, {Mohammad}, {Mosser}, {Mueller}, {Muna},
  {Mu{\~n}oz-Guti{\'e}rrez}, {Myers}, {Nadathur}, {Nair}, {Nandra}, {Correa do
  Nascimento}, {Nevin}, {Newman}, {Nidever}, {Nitschelm}, {Noterdaeme},
  {O'Connell}, {Olmstead}, {Oravetz}, {Oravetz}, {Osorio}, {Pace}, {Padilla},
  {Palanque-Delabrouille}, {Palicio}, {Pan}, {Pan}, {Parker}, {Paviot},
  {Peirani}, {Ram{\'r}ez}, {Penny}, {Percival}, {Perez-Fournon},
  {P{\'e}rez-R{\`a}fols}, {Petitjean}, {Pieri}, {Pinsonneault}, {Poovelil},
  {Povick}, {Prakash}, {Price-Whelan}, {Raddick}, {Raichoor}, {Ray}, {Rembold},
  {Rezaie}, {Riffel}, {Riffel}, {Rix}, {Robin}, {Roman-Lopes},
  {Rom{\'a}n-Z{\'u}{\~n}iga}, {Rose}, {Ross}, {Rossi}, {Rowlands}, {Rubin},
  {Salvato}, {S{\'a}nchez}, {S{\'a}nchez-Menguiano}, {S{\'a}nchez-Gallego},
  {Sayres}, {Schaefer}, {Schiavon}, {Schimoia}, {Schlafly}, {Schlegel},
  {Schneider}, {Schultheis}, {Schwope}, {Seo}, {Serenelli}, {Shafieloo},
  {Shamsi}, {Shao}, {Shen}, {Shetrone}, {Shirley}, {Silva Aguirre}, {Simon},
  {Skrutskie}, {Slosar}, {Smethurst}, {Sobeck}, {Sodi}, {Souto}, {Stark},
  {Stassun}, {Steinmetz}, {Stello}, {Stermer}, {Storchi-Bergmann},
  {Streblyanska}, {Stringfellow}, {Stutz}, {Su{\'a}rez}, {Sun},
  {Taghizadeh-Popp}, {Talbot}, {Tayar}, {Thakar}, {Theriault}, {Thomas},
  {Thomas}, {Tinker}, {Tojeiro}, {Toledo}, {Tremonti}, {Troup}, {Tuttle},
  {Unda-Sanzana}, {Valentini}, {Vargas-Gonz{\'a}lez}, {Vargas-Maga{\~n}a},
  {V{\'a}zquez-Mata}, {Vivek}, {Wake}, {Wang}, {Weaver}, {Weijmans}, {Wild},
  {Wilson}, {Wilson}, {Wolthuis}, {Wood-Vasey}, {Yan}, {Yang}, {Y{\`e}che},
  {Zamora}, {Zarrouk}, {Zasowski}, {Zhang}, {Zhao}, {Zhao}, {Zheng}, {Zheng},
  {Zhu}, \& {Zou}}]{2020ApJS..249....3A}
{Ahumada}, R., {Allende Prieto}, C., {Almeida}, A., {et~al.} 2020, \apjs, 249,
  3

\bibitem[{{Akhlaghi} \& {Ichikawa}(2015)}]{2015ApJS..220....1A}
{Akhlaghi}, M. \& {Ichikawa}, T. 2015, \apjs, 220, 1

\bibitem[{{Argudo-Fern{\'a}ndez} {et~al.}(2017){Argudo-Fern{\'a}ndez}, {Duarte
  Puertas}, {Ruiz}, {Sabater}, {Verley}, \& {Bergond}}]{2017PASP..129e8005A}
{Argudo-Fern{\'a}ndez}, M., {Duarte Puertas}, S., {Ruiz}, J.~E., {et~al.} 2017,
  \pasp, 129, 058005

\bibitem[{{Argudo-Fern{\'a}ndez} {et~al.}(2015){Argudo-Fern{\'a}ndez},
  {Verley}, {Bergond}, {Duarte Puertas}, {Ramos Carmona}, {Sabater},
  {Fern{\'a}ndez Lorenzo}, {Espada}, {Sulentic}, {Ruiz}, \&
  {Leon}}]{2015A&A...578A.110A}
{Argudo-Fern{\'a}ndez}, M., {Verley}, S., {Bergond}, G., {et~al.} 2015, \aap,
  578, A110

\bibitem[{{Astropy Collaboration} {et~al.}(2018){Astropy Collaboration},
  {Price-Whelan}, {Sip{\H{o}}cz}, {G{\"u}nther}, {Lim}, {Crawford}, {Conseil},
  {Shupe}, {Craig}, {Dencheva}, {Ginsburg}, {VanderPlas}, {Bradley},
  {P{\'e}rez-Su{\'a}rez}, {de Val-Borro}, {Aldcroft}, {Cruz}, {Robitaille},
  {Tollerud}, {Ardelean}, {Babej}, {Bach}, {Bachetti}, {Bakanov}, {Bamford},
  {Barentsen}, {Barmby}, {Baumbach}, {Berry}, {Biscani}, {Boquien}, {Bostroem},
  {Bouma}, {Brammer}, {Bray}, {Breytenbach}, {Buddelmeijer}, {Burke},
  {Calderone}, {Cano Rodr{\'\i}guez}, {Cara}, {Cardoso}, {Cheedella}, {Copin},
  {Corrales}, {Crichton}, {D'Avella}, {Deil}, {Depagne}, {Dietrich}, {Donath},
  {Droettboom}, {Earl}, {Erben}, {Fabbro}, {Ferreira}, {Finethy}, {Fox},
  {Garrison}, {Gibbons}, {Goldstein}, {Gommers}, {Greco}, {Greenfield},
  {Groener}, {Grollier}, {Hagen}, {Hirst}, {Homeier}, {Horton}, {Hosseinzadeh},
  {Hu}, {Hunkeler}, {Ivezi{\'c}}, {Jain}, {Jenness}, {Kanarek}, {Kendrew},
  {Kern}, {Kerzendorf}, {Khvalko}, {King}, {Kirkby}, {Kulkarni}, {Kumar},
  {Lee}, {Lenz}, {Littlefair}, {Ma}, {Macleod}, {Mastropietro}, {McCully},
  {Montagnac}, {Morris}, {Mueller}, {Mumford}, {Muna}, {Murphy}, {Nelson},
  {Nguyen}, {Ninan}, {N{\"o}the}, {Ogaz}, {Oh}, {Parejko}, {Parley}, {Pascual},
  {Patil}, {Patil}, {Plunkett}, {Prochaska}, {Rastogi}, {Reddy Janga},
  {Sabater}, {Sakurikar}, {Seifert}, {Sherbert}, {Sherwood-Taylor}, {Shih},
  {Sick}, {Silbiger}, {Singanamalla}, {Singer}, {Sladen}, {Sooley},
  {Sornarajah}, {Streicher}, {Teuben}, {Thomas}, {Tremblay}, {Turner},
  {Terr{\'o}n}, {van Kerkwijk}, {de la Vega}, {Watkins}, {Weaver}, {Whitmore},
  {Woillez}, {Zabalza}, \& {Astropy Contributors}}]{astropy2018}
{Astropy Collaboration}, {Price-Whelan}, A.~M., {Sip{\H{o}}cz}, B.~M., {et~al.}
  2018, \aj, 156, 123

\bibitem[{{Astropy Collaboration} {et~al.}(2013){Astropy Collaboration},
  {Robitaille}, {Tollerud}, {Greenfield}, {Droettboom}, {Bray}, {Aldcroft},
  {Davis}, {Ginsburg}, {Price-Whelan}, {Kerzendorf}, {Conley}, {Crighton},
  {Barbary}, {Muna}, {Ferguson}, {Grollier}, {Parikh}, {Nair}, {Unther},
  {Deil}, {Woillez}, {Conseil}, {Kramer}, {Turner}, {Singer}, {Fox}, {Weaver},
  {Zabalza}, {Edwards}, {Azalee Bostroem}, {Burke}, {Casey}, {Crawford},
  {Dencheva}, {Ely}, {Jenness}, {Labrie}, {Lim}, {Pierfederici}, {Pontzen},
  {Ptak}, {Refsdal}, {Servillat}, \& {Streicher}}]{astropy}
{Astropy Collaboration}, {Robitaille}, T.~P., {Tollerud}, E.~J., {et~al.} 2013,
  \aap, 558, A33

\bibitem[{Behnel {et~al.}(2011)Behnel, Bradshaw, Citro, Dalcin, Seljebotn, \&
  Smith}]{behnel2011cython}
Behnel, S., Bradshaw, R., Citro, C., {et~al.} 2011, Computing in Science \&
  Engineering, 13, 31

\bibitem[{{Bertin}(2006)}]{2006ASPC..351..112B}
{Bertin}, E. 2006, in Astronomical Society of the Pacific Conference Series,
  Vol. 351, Astronomical Data Analysis Software and Systems XV, ed.
  C.~{Gabriel}, C.~{Arviset}, D.~{Ponz}, \& S.~{Enrique}, 112

\bibitem[{{Bertin} \& {Arnouts}(1996)}]{1996A&AS..117..393B}
{Bertin}, E. \& {Arnouts}, S. 1996, \aaps, 117, 393

\bibitem[{{Beygu} {et~al.}(2016){Beygu}, {Kreckel}, {van der Hulst}, {Jarrett},
  {Peletier}, {van de Weygaert}, {van Gorkom}, \&
  {Aragon-Calvo}}]{2016MNRAS.458..394B}
{Beygu}, B., {Kreckel}, K., {van der Hulst}, J.~M., {et~al.} 2016, \mnras, 458,
  394

\bibitem[{{Beygu} {et~al.}(2017){Beygu}, {Peletier}, {van der Hulst},
  {Jarrett}, {Kreckel}, {van de Weygaert}, {van Gorkom}, \&
  {Aragon-Calvo}}]{2017MNRAS.464..666B}
{Beygu}, B., {Peletier}, R.~F., {van der Hulst}, J.~M., {et~al.} 2017, \mnras,
  464, 666

\bibitem[{{Brassington} {et~al.}(2015){Brassington}, {Zezas}, {Ashby}, {Lanz},
  {Smith}, {Willner}, \& {Klein}}]{2015ApJS..218....6B}
{Brassington}, N.~J., {Zezas}, A., {Ashby}, M.~L.~N., {et~al.} 2015, \apjs,
  218, 6

\bibitem[{{Bundy} {et~al.}(2015){Bundy}, {Bershady}, {Law}, {Yan}, {Drory},
  {MacDonald}, {Wake}, {Cherinka}, {S{\'a}nchez-Gallego}, {Weijmans}, {Thomas},
  {Tremonti}, {Masters}, {Coccato}, {Diamond-Stanic}, {Arag{\'o}n-Salamanca},
  {Avila-Reese}, {Badenes}, {Falc{\'o}n-Barroso}, {Belfiore}, {Bizyaev},
  {Blanc}, {Bland-Hawthorn}, {Blanton}, {Brownstein}, {Byler}, {Cappellari},
  {Conroy}, {Dutton}, {Emsellem}, {Etherington}, {Frinchaboy}, {Fu}, {Gunn},
  {Harding}, {Johnston}, {Kauffmann}, {Kinemuchi}, {Klaene}, {Knapen},
  {Leauthaud}, {Li}, {Lin}, {Maiolino}, {Malanushenko}, {Malanushenko}, {Mao},
  {Maraston}, {McDermid}, {Merrifield}, {Nichol}, {Oravetz}, {Pan}, {Parejko},
  {Sanchez}, {Schlegel}, {Simmons}, {Steele}, {Steinmetz}, {Thanjavur},
  {Thompson}, {Tinker}, {van den Bosch}, {Westfall}, {Wilkinson}, {Wright},
  {Xiao}, \& {Zhang}}]{2015ApJ...798....7B}
{Bundy}, K., {Bershady}, M.~A., {Law}, D.~R., {et~al.} 2015, \apj, 798, 7

\bibitem[{{Cappellari} \& {Copin}(2003)}]{2003MNRAS.342..345C}
{Cappellari}, M. \& {Copin}, Y. 2003, \mnras, 342, 345

\bibitem[{{Cappellari} \& {Emsellem}(2004)}]{2004PASP..116..138C}
{Cappellari}, M. \& {Emsellem}, E. 2004, \pasp, 116, 138

\bibitem[{{Cardelli} {et~al.}(1989){Cardelli}, {Clayton}, \&
  {Mathis}}]{1989ApJ...345..245C}
{Cardelli}, J.~A., {Clayton}, G.~C., \& {Mathis}, J.~S. 1989, \apj, 345, 245

\bibitem[{{Cautun} {et~al.}(2014){Cautun}, {van de Weygaert}, {Jones}, \&
  {Frenk}}]{2014MNRAS.441.2923C}
{Cautun}, M., {van de Weygaert}, R., {Jones}, B. J.~T., \& {Frenk}, C.~S. 2014,
  \mnras, 441, 2923

\bibitem[{{Cid Fernandes} {et~al.}(2013){Cid Fernandes}, {P{\'e}rez},
  {Garc{\'\i}a Benito}, {Gonz{\'a}lez Delgado}, {de Amorim}, {S{\'a}nchez},
  {Husemann}, {Falc{\'o}n Barroso}, {S{\'a}nchez-Bl{\'a}zquez}, {Walcher}, \&
  {Mast}}]{2013A&A...557A..86C}
{Cid Fernandes}, R., {P{\'e}rez}, E., {Garc{\'\i}a Benito}, R., {et~al.} 2013,
  \aap, 557, A86

\bibitem[{{Colberg} {et~al.}(2008){Colberg}, {Pearce}, {Foster}, {Platen},
  {Brunino}, {Neyrinck}, {Basilakos}, {Fairall}, {Feldman}, {Gottl{\"o}ber},
  {Hahn}, {Hoyle}, {M{\"u}ller}, {Nelson}, {Plionis}, {Porciani}, {Shandarin},
  {Vogeley}, \& {van de Weygaert}}]{2008MNRAS.387..933C}
{Colberg}, J.~M., {Pearce}, F., {Foster}, C., {et~al.} 2008, \mnras, 387, 933

\bibitem[{{Conrado} {et~al.}(2024){Conrado}, {Gonz{\'a}lez Delgado},
  {Garc{\'\i}a-Benito}, {P{\'e}rez}, {Verley}, {Ruiz-Lara},
  {S{\'a}nchez-Menguiano}, {Duarte Puertas}, {Jim{\'e}nez},
  {Dom{\'\i}nguez-G{\'o}mez}, {Espada}, {Argudo-Fern{\'a}ndez},
  {Alc{\'a}zar-Laynez}, {Bl{\'a}zquez-Calero}, {Bidaran}, {Zurita}, {Peletier},
  {Torres-R{\'\i}os}, {Florido}, {Rodr{\'\i}guez Mart{\'\i}nez}, {del
  Moral-Castro}, {van de Weygaert}, {Falc{\'o}n-Barroso}, {Lugo-Aranda},
  {S{\'a}nchez}, {van der Hulst}, {Courtois}, {Ferr{\'e}-Mateu},
  {S{\'a}nchez-Bl{\'a}zquez}, {Rom{\'a}n}, \& {Aceituno}}]{2024arXiv240410823C}
{Conrado}, A.~M., {Gonz{\'a}lez Delgado}, R.~M., {Garc{\'\i}a-Benito}, R.,
  {et~al.} 2024, arXiv e-prints, arXiv:2404.10823

\bibitem[{{Constantin} {et~al.}(2008){Constantin}, {Hoyle}, \&
  {Vogeley}}]{2008ApJ...673..715C}
{Constantin}, A., {Hoyle}, F., \& {Vogeley}, M.~S. 2008, \apj, 673, 715

\bibitem[{{Debattista} {et~al.}(2006){Debattista}, {Mayer}, {Carollo}, {Moore},
  {Wadsley}, \& {Quinn}}]{2006ApJ...645..209D}
{Debattista}, V.~P., {Mayer}, L., {Carollo}, C.~M., {et~al.} 2006, \apj, 645,
  209

\bibitem[{{Dekel} \& {Cox}(2006)}]{2006MNRAS.370.1445D}
{Dekel}, A. \& {Cox}, T.~J. 2006, \mnras, 370, 1445

\bibitem[{{del Moral-Castro} {et~al.}(2019){del Moral-Castro},
  {Garc{\'\i}a-Lorenzo}, {Ramos Almeida}, {Ruiz-Lara}, {Falc{\'o}n-Barroso},
  {S{\'a}nchez}, {S{\'a}nchez-Bl{\'a}zquez}, {M{\'a}rquez}, \&
  {Masegosa}}]{2019MNRAS.485.3794D}
{del Moral-Castro}, I., {Garc{\'\i}a-Lorenzo}, B., {Ramos Almeida}, C.,
  {et~al.} 2019, \mnras, 485, 3794

\bibitem[{{del Moral-Castro} {et~al.}(2020){del Moral-Castro},
  {Garc{\'\i}a-Lorenzo}, {Ramos Almeida}, {Ruiz-Lara}, {Falc{\'o}n-Barroso},
  {S{\'a}nchez}, {S{\'a}nchez-Bl{\'a}zquez}, {M{\'a}rquez}, \&
  {Masegosa}}]{2020A&A...639L...9D}
{del Moral-Castro}, I., {Garc{\'\i}a-Lorenzo}, B., {Ramos Almeida}, C.,
  {et~al.} 2020, \aap, 639, L9

\bibitem[{{Dey} {et~al.}(2019){Dey}, {Schlegel}, {Lang}, {Blum}, {Burleigh},
  {Fan}, {Findlay}, {Finkbeiner}, {Herrera}, {Juneau}, {Landriau}, {Levi},
  {McGreer}, {Meisner}, {Myers}, {Moustakas}, {Nugent}, {Patej}, {Schlafly},
  {Walker}, {Valdes}, {Weaver}, {Y{\`e}che}, {Zou}, {Zhou}, {Abareshi},
  {Abbott}, {Abolfathi}, {Aguilera}, {Alam}, {Allen}, {Alvarez}, {Annis},
  {Ansarinejad}, {Aubert}, {Beechert}, {Bell}, {BenZvi}, {Beutler}, {Bielby},
  {Bolton}, {Brice{\~n}o}, {Buckley-Geer}, {Butler}, {Calamida}, {Carlberg},
  {Carter}, {Casas}, {Castander}, {Choi}, {Comparat}, {Cukanovaite}, {Delubac},
  {DeVries}, {Dey}, {Dhungana}, {Dickinson}, {Ding}, {Donaldson}, {Duan},
  {Duckworth}, {Eftekharzadeh}, {Eisenstein}, {Etourneau}, {Fagrelius},
  {Farihi}, {Fitzpatrick}, {Font-Ribera}, {Fulmer}, {G{\"a}nsicke},
  {Gaztanaga}, {George}, {Gerdes}, {Gontcho}, {Gorgoni}, {Green}, {Guy},
  {Harmer}, {Hernandez}, {Honscheid}, {Huang}, {James}, {Jannuzi}, {Jiang},
  {Joyce}, {Karcher}, {Karkar}, {Kehoe}, {Kneib}, {Kueter-Young}, {Lan},
  {Lauer}, {Le Guillou}, {Le Van Suu}, {Lee}, {Lesser}, {Perreault Levasseur},
  {Li}, {Mann}, {Marshall}, {Mart{\'\i}nez-V{\'a}zquez}, {Martini}, {du Mas des
  Bourboux}, {McManus}, {Meier}, {M{\'e}nard}, {Metcalfe},
  {Mu{\~n}oz-Guti{\'e}rrez}, {Najita}, {Napier}, {Narayan}, {Newman}, {Nie},
  {Nord}, {Norman}, {Olsen}, {Paat}, {Palanque-Delabrouille}, {Peng},
  {Poppett}, {Poremba}, {Prakash}, {Rabinowitz}, {Raichoor}, {Rezaie},
  {Robertson}, {Roe}, {Ross}, {Ross}, {Rudnick}, {Safonova}, {Saha},
  {S{\'a}nchez}, {Savary}, {Schweiker}, {Scott}, {Seo}, {Shan}, {Silva},
  {Slepian}, {Soto}, {Sprayberry}, {Staten}, {Stillman}, {Stupak}, {Summers},
  {Sien Tie}, {Tirado}, {Vargas-Maga{\~n}a}, {Vivas}, {Wechsler}, {Williams},
  {Yang}, {Yang}, {Yapici}, {Zaritsky}, {Zenteno}, {Zhang}, {Zhang}, {Zhou}, \&
  {Zhou}}]{2019AJ....157..168D}
{Dey}, A., {Schlegel}, D.~J., {Lang}, D., {et~al.} 2019, \aj, 157, 168

\bibitem[{{Di Matteo} {et~al.}(2005){Di Matteo}, {Springel}, \&
  {Hernquist}}]{2005Natur.433..604D}
{Di Matteo}, T., {Springel}, V., \& {Hernquist}, L. 2005, \nat, 433, 604

\bibitem[{{Dom{\'\i}nguez-G{\'o}mez} {et~al.}(2022){Dom{\'\i}nguez-G{\'o}mez},
  {Lisenfeld}, {P{\'e}rez}, {L{\'o}pez-S{\'a}nchez}, {Duarte Puertas},
  {Falc{\'o}n-Barroso}, {Kreckel}, {Peletier}, {Ruiz-Lara}, {van de Weygaert},
  {van der Hulst}, \& {Verley}}]{2022A&A...658A.124D}
{Dom{\'\i}nguez-G{\'o}mez}, J., {Lisenfeld}, U., {P{\'e}rez}, I., {et~al.}
  2022, \aap, 658, A124

\bibitem[{{Dom{\'\i}nguez-G{\'o}mez}
  {et~al.}(2023{\natexlab{a}}){Dom{\'\i}nguez-G{\'o}mez}, {P{\'e}rez},
  {Ruiz-Lara}, {Peletier}, {S{\'a}nchez-Bl{\'a}zquez}, {Lisenfeld}, {Bidaran},
  {Falc{\'o}n-Barroso}, {Alc{\'a}zar-Laynez}, {Argudo-Fern{\'a}ndez},
  {Bl{\'a}zquez-Calero}, {Courtois}, {Duarte Puertas}, {Espada}, {Florido},
  {Garc{\'\i}a-Benito}, {Jim{\'e}nez}, {Kreckel}, {Rela{\~n}o},
  {S{\'a}nchez-Menguiano}, {van der Hulst}, {van de Weygaert}, {Verley}, \&
  {Zurita}}]{2023arXiv231011412D}
{Dom{\'\i}nguez-G{\'o}mez}, J., {P{\'e}rez}, I., {Ruiz-Lara}, T., {et~al.}
  2023{\natexlab{a}}, arXiv e-prints, arXiv:2310.11412

\bibitem[{{Dom{\'\i}nguez-G{\'o}mez}
  {et~al.}(2023{\natexlab{b}}){Dom{\'\i}nguez-G{\'o}mez}, {P{\'e}rez},
  {Ruiz-Lara}, {Peletier}, {S{\'a}nchez-Bl{\'a}zquez}, {Lisenfeld},
  {Falc{\'o}n-Barroso}, {Alc{\'a}zar-Laynez}, {Argudo-Fern{\'a}ndez},
  {Bl{\'a}zquez-Calero}, {Courtois}, {Duarte Puertas}, {Espada}, {Florido},
  {Garc{\'\i}a-Benito}, {Jim{\'e}nez}, {Kreckel}, {Rela{\~n}o},
  {S{\'a}nchez-Menguiano}, {van der Hulst}, {van de Weygaert}, {Verley}, \&
  {Zurita}}]{2023Natur.619..269D}
{Dom{\'\i}nguez-G{\'o}mez}, J., {P{\'e}rez}, I., {Ruiz-Lara}, T., {et~al.}
  2023{\natexlab{b}}, \nat, 619, 269

\bibitem[{{Douglass} {et~al.}(2019){Douglass}, {Smith}, \&
  {Demina}}]{2019ApJ...886..153D}
{Douglass}, K.~A., {Smith}, J.~A., \& {Demina}, R. 2019, \apj, 886, 153

\bibitem[{{Douglass} {et~al.}(2022){Douglass}, {Veyrat}, \&
  {BenZvi}}]{2022arXiv220201226D}
{Douglass}, K.~A., {Veyrat}, D., \& {BenZvi}, S. 2022, arXiv e-prints,
  arXiv:2202.01226

\bibitem[{{Dressler}(1980)}]{1980ApJ...236..351D}
{Dressler}, A. 1980, \apj, 236, 351

\bibitem[{{El-Ad} \& {Piran}(1997)}]{1997ApJ...491..421E}
{El-Ad}, H. \& {Piran}, T. 1997, \apj, 491, 421

\bibitem[{{Falc{\'o}n-Barroso} {et~al.}(2006){Falc{\'o}n-Barroso}, {Bacon},
  {Bureau}, {Cappellari}, {Davies}, {de Zeeuw}, {Emsellem}, {Fathi},
  {Krajnovi{\'c}}, {Kuntschner}, {McDermid}, {Peletier}, \&
  {Sarzi}}]{2006MNRAS.369..529F}
{Falc{\'o}n-Barroso}, J., {Bacon}, R., {Bureau}, M., {et~al.} 2006, \mnras,
  369, 529

\bibitem[{{Falc{\'o}n-Barroso} {et~al.}(2011){Falc{\'o}n-Barroso},
  {S{\'a}nchez-Bl{\'a}zquez}, {Vazdekis}, {Ricciardelli}, {Cardiel}, {Cenarro},
  {Gorgas}, \& {Peletier}}]{2011A&A...532A..95F}
{Falc{\'o}n-Barroso}, J., {S{\'a}nchez-Bl{\'a}zquez}, P., {Vazdekis}, A.,
  {et~al.} 2011, \aap, 532, A95

\bibitem[{{Falc{\'o}n-Barroso} {et~al.}(2019){Falc{\'o}n-Barroso}, {van de
  Ven}, {Lyubenova}, {Mendez-Abreu}, {Aguerri}, {Garc{\'\i}a-Lorenzo},
  {Bekerait{\'e}}, {S{\'a}nchez}, {Husemann}, {Garc{\'\i}a-Benito},
  {Gonz{\'a}lez Delgado}, {Mast}, {Walcher}, {Zibetti}, {Zhu},
  {Barrera-Ballesteros}, {Galbany}, {S{\'a}nchez-Bl{\'a}zquez}, {Singh}, {van
  den Bosch}, {Wild}, {Bland-Hawthorn}, {Cid Fernandes}, {de
  Lorenzo-C{\'a}ceres}, {Gallazzi}, {Marino}, {M{\'a}rquez}, {Peletier},
  {P{\'e}rez}, {P{\'e}rez}, {Roth}, {Rosales-Ortega}, {Ruiz-Lara}, {Wisotzki},
  \& {Ziegler}}]{2019A&A...632A..59F}
{Falc{\'o}n-Barroso}, J., {van de Ven}, G., {Lyubenova}, M., {et~al.} 2019,
  \aap, 632, A59

\bibitem[{{Garc{\'\i}a-Benito} {et~al.}(2017){Garc{\'\i}a-Benito},
  {Gonz{\'a}lez Delgado}, {P{\'e}rez}, {Cid Fernandes}, {Cortijo-Ferrero},
  {L{\'o}pez Fern{\'a}ndez}, {de Amorim}, {Lacerda}, {Vale Asari}, \&
  {S{\'a}nchez}}]{2017A&A...608A..27G}
{Garc{\'\i}a-Benito}, R., {Gonz{\'a}lez Delgado}, R.~M., {P{\'e}rez}, E.,
  {et~al.} 2017, \aap, 608, A27

\bibitem[{{Garc{\'\i}a-Benito} \&
  {P{\'e}rez-Montero}(2012)}]{2012MNRAS.423..406G}
{Garc{\'\i}a-Benito}, R. \& {P{\'e}rez-Montero}, E. 2012, \mnras, 423, 406

\bibitem[{{Hopkins} {et~al.}(2006){Hopkins}, {Somerville}, {Hernquist}, {Cox},
  {Robertson}, \& {Li}}]{2006ApJ...652..864H}
{Hopkins}, P.~F., {Somerville}, R.~S., {Hernquist}, L., {et~al.} 2006, \apj,
  652, 864

\bibitem[{{Hoyle} \& {Vogeley}(2002)}]{2002ApJ...566..641H}
{Hoyle}, F. \& {Vogeley}, M.~S. 2002, \apj, 566, 641

\bibitem[{{Hoyle} \& {Vogeley}(2004)}]{2004ApJ...607..751H}
{Hoyle}, F. \& {Vogeley}, M.~S. 2004, \apj, 607, 751

\bibitem[{{Husemann} {et~al.}(2013){Husemann}, {Jahnke}, {S{\'a}nchez},
  {Barrado}, {Bekerait{\.{e}}}, {Bomans}, {Castillo-Morales},
  {Catal{\'a}n-Torrecilla}, {Cid Fernandes}, {Falc{\'o}n-Barroso},
  {Garc{\'\i}a-Benito}, {Gonz{\'a}lez Delgado}, {Iglesias-P{\'a}ramo},
  {Johnson}, {Kupko}, {L{\'o}pez-Fernandez}, {Lyubenova}, {Marino}, {Mast},
  {Miskolczi}, {Monreal-Ibero}, {Gil de Paz}, {P{\'e}rez}, {P{\'e}rez},
  {Rosales-Ortega}, {Ruiz-Lara}, {Schilling}, {van de Ven}, {Walcher}, {Alves},
  {de Amorim}, {Backsmann}, {Barrera-Ballesteros}, {Bland-Hawthorn}, {Cortijo},
  {Dettmar}, {Demleitner}, {D{\'\i}az}, {Enke}, {Florido}, {Flores}, {Galbany},
  {Gallazzi}, {Garc{\'\i}a-Lorenzo}, {Gomes}, {Gruel}, {Haines}, {Holmes},
  {Jungwiert}, {Kalinova}, {Kehrig}, {Kennicutt}, {Klar}, {Lehnert},
  {L{\'o}pez-S{\'a}nchez}, {de Lorenzo-C{\'a}ceres}, {M{\'a}rmol-Queralt{\'o}},
  {M{\'a}rquez}, {Mendez-Abreu}, {Moll{\'a}}, {del Olmo}, {Meidt}, {Papaderos},
  {Puschnig}, {Quirrenbach}, {Roth}, {S{\'a}nchez-Bl{\'a}zquez}, {Spekkens},
  {Singh}, {Stanishev}, {Trager}, {Vilchez}, {Wild}, {Wisotzki}, {Zibetti}, \&
  {Ziegler}}]{2013A&A...549A..87H}
{Husemann}, B., {Jahnke}, K., {S{\'a}nchez}, S.~F., {et~al.} 2013, \aap, 549,
  A87

\bibitem[{{Husemann} {et~al.}(2012){Husemann}, {Kamann}, {Sandin},
  {S{\'a}nchez}, {Garc{\'\i}a-Benito}, \& {Mast}}]{2012A&A...545A.137H}
{Husemann}, B., {Kamann}, S., {Sandin}, C., {et~al.} 2012, \aap, 545, A137

\bibitem[{{Jedrzejewski}(1987)}]{Jedrzejewski1987}
{Jedrzejewski}, R.~I. 1987, \mnras, 226, 747

\bibitem[{{Kelz} {et~al.}(2006){Kelz}, {Verheijen}, {Roth}, {Bauer}, {Becker},
  {Paschke}, {Popow}, {S{\'a}nchez}, \& {Laux}}]{2006PASP..118..129K}
{Kelz}, A., {Verheijen}, M. A.~W., {Roth}, M.~M., {et~al.} 2006, \pasp, 118,
  129

\bibitem[{{Klypin} {et~al.}(2009){Klypin}, {Valenzuela}, {Col{\'\i}n}, \&
  {Quinn}}]{2009MNRAS.398.1027K}
{Klypin}, A., {Valenzuela}, O., {Col{\'\i}n}, P., \& {Quinn}, T. 2009, \mnras,
  398, 1027

\bibitem[{{Knapen} \& {Cisternas}(2015)}]{2015ApJ...807L..16K}
{Knapen}, J.~H. \& {Cisternas}, M. 2015, \apjl, 807, L16

\bibitem[{{Kreckel} {et~al.}(2015){Kreckel}, {Croxall}, {Groves}, {van de
  Weygaert}, \& {Pogge}}]{2015ApJ...798L..15K}
{Kreckel}, K., {Croxall}, K., {Groves}, B., {van de Weygaert}, R., \& {Pogge},
  R.~W. 2015, \apjl, 798, L15

\bibitem[{{Kroupa}(2001)}]{2001MNRAS.322..231K}
{Kroupa}, P. 2001, \mnras, 322, 231

\bibitem[{{Lacerda} {et~al.}(2022){Lacerda}, {S{\'a}nchez},
  {Mej{\'\i}a-Narv{\'a}ez}, {Camps-Fari{\~n}a}, {Espinosa-Ponce},
  {Barrera-Ballesteros}, {Ibarra-Medel}, \&
  {Lugo-Aranda}}]{2022NewA...9701895L}
{Lacerda}, E. A.~D., {S{\'a}nchez}, S.~F., {Mej{\'\i}a-Narv{\'a}ez}, A.,
  {et~al.} 2022, \na, 97, 101895

\bibitem[{{Lackner} {et~al.}(2012){Lackner}, {Cen}, {Ostriker}, \&
  {Joung}}]{2012MNRAS.425..641L}
{Lackner}, C.~N., {Cen}, R., {Ostriker}, J.~P., \& {Joung}, M.~R. 2012, \mnras,
  425, 641

\bibitem[{{Lang} {et~al.}(2010){Lang}, {Hogg}, {Mierle}, {Blanton}, \&
  {Roweis}}]{2010AJ....139.1782L}
{Lang}, D., {Hogg}, D.~W., {Mierle}, K., {Blanton}, M., \& {Roweis}, S. 2010,
  \aj, 139, 1782

\bibitem[{{Larson} \& {Tinsley}(1978)}]{1978ApJ...219...46L}
{Larson}, R.~B. \& {Tinsley}, B.~M. 1978, \apj, 219, 46

\bibitem[{{Libeskind} {et~al.}(2018){Libeskind}, {van de Weygaert}, {Cautun},
  {Falck}, {Tempel}, {Abel}, {Alpaslan}, {Arag{\'o}n-Calvo}, {Forero-Romero},
  {Gonzalez}, {Gottl{\"o}ber}, {Hahn}, {Hellwing}, {Hoffman}, {Jones},
  {Kitaura}, {Knebe}, {Manti}, {Neyrinck}, {Nuza}, {Padilla}, {Platen},
  {Ramachandra}, {Robotham}, {Saar}, {Shandarin}, {Steinmetz}, {Stoica},
  {Sousbie}, \& {Yepes}}]{2018MNRAS.473.1195L}
{Libeskind}, N.~I., {van de Weygaert}, R., {Cautun}, M., {et~al.} 2018, \mnras,
  473, 1195

\bibitem[{{Maiolino} {et~al.}(2012){Maiolino}, {Gallerani}, {Neri}, {Cicone},
  {Ferrara}, {Genzel}, {Lutz}, {Sturm}, {Tacconi}, {Walter}, {Feruglio},
  {Fiore}, \& {Piconcelli}}]{2012MNRAS.425L..66M}
{Maiolino}, R., {Gallerani}, S., {Neri}, R., {et~al.} 2012, \mnras, 425, L66

\bibitem[{{Makarov} {et~al.}(2014){Makarov}, {Prugniel}, {Terekhova},
  {Courtois}, \& {Vauglin}}]{2014A&A...570A..13M}
{Makarov}, D., {Prugniel}, P., {Terekhova}, N., {Courtois}, H., \& {Vauglin},
  I. 2014, \aap, 570, A13

\bibitem[{{Marino} {et~al.}(2013){Marino}, {Rosales-Ortega}, {S{\'a}nchez},
  {Gil de Paz}, {V{\'\i}lchez}, {Miralles-Caballero}, {Kehrig},
  {P{\'e}rez-Montero}, {Stanishev}, {Iglesias-P{\'a}ramo}, {D{\'\i}az},
  {Castillo-Morales}, {Kennicutt}, {L{\'o}pez-S{\'a}nchez}, {Galbany},
  {Garc{\'\i}a-Benito}, {Mast}, {Mendez-Abreu}, {Monreal-Ibero}, {Husemann},
  {Walcher}, {Garc{\'\i}a-Lorenzo}, {Masegosa}, {Del Olmo Orozco},
  {Mour{\~a}o}, {Ziegler}, {Moll{\'a}}, {Papaderos},
  {S{\'a}nchez-Bl{\'a}zquez}, {Gonz{\'a}lez Delgado}, {Falc{\'o}n-Barroso},
  {Roth}, {van de Ven}, \& {CALIFA Team}}]{2013A&A...559A.114M}
{Marino}, R.~A., {Rosales-Ortega}, F.~F., {S{\'a}nchez}, S.~F., {et~al.} 2013,
  \aap, 559, A114

\bibitem[{{Moster} {et~al.}(2013){Moster}, {Naab}, \&
  {White}}]{2013MNRAS.428.3121M}
{Moster}, B.~P., {Naab}, T., \& {White}, S. D.~M. 2013, \mnras, 428, 3121

\bibitem[{{Nadathur} \& {Hotchkiss}(2014)}]{2014MNRAS.440.1248N}
{Nadathur}, S. \& {Hotchkiss}, S. 2014, \mnras, 440, 1248

\bibitem[{{Neyrinck}(2008)}]{2008MNRAS.386.2101N}
{Neyrinck}, M.~C. 2008, \mnras, 386, 2101

\bibitem[{{Nikolic} {et~al.}(2004){Nikolic}, {Cullen}, \&
  {Alexander}}]{2004MNRAS.355..874N}
{Nikolic}, B., {Cullen}, H., \& {Alexander}, P. 2004, \mnras, 355, 874

\bibitem[{{Pan} {et~al.}(2012){Pan}, {Vogeley}, {Hoyle}, {Choi}, \&
  {Park}}]{2012MNRAS.421..926P}
{Pan}, D.~C., {Vogeley}, M.~S., {Hoyle}, F., {Choi}, Y.-Y., \& {Park}, C. 2012,
  \mnras, 421, 926

\bibitem[{{Park} {et~al.}(2007){Park}, {Choi}, {Vogeley}, {Gott}, {Blanton}, \&
  {SDSS Collaboration}}]{2007ApJ...658..898P}
{Park}, C., {Choi}, Y.-Y., {Vogeley}, M.~S., {et~al.} 2007, \apj, 658, 898

\bibitem[{{P{\'e}rez} {et~al.}(2017){P{\'e}rez}, {Mart{\'\i}nez-Valpuesta},
  {Ruiz-Lara}, {de Lorenzo-Caceres}, {Falc{\'o}n-Barroso}, {Florido},
  {Gonz{\'a}lez Delgado}, {Lyubenova}, {Marino}, {S{\'a}nchez},
  {S{\'a}nchez-Bl{\'a}zquez}, {van de Ven}, \& {Zurita}}]{2017MNRAS.470L.122P}
{P{\'e}rez}, I., {Mart{\'\i}nez-Valpuesta}, I., {Ruiz-Lara}, T., {et~al.} 2017,
  \mnras, 470, L122

\bibitem[{{Pietrinferni} {et~al.}(2004){Pietrinferni}, {Cassisi}, {Salaris}, \&
  {Castelli}}]{2004ApJ...612..168P}
{Pietrinferni}, A., {Cassisi}, S., {Salaris}, M., \& {Castelli}, F. 2004, \apj,
  612, 168

\bibitem[{{Platen} {et~al.}(2008){Platen}, {van de Weygaert}, \&
  {Jones}}]{2008MNRAS.387..128P}
{Platen}, E., {van de Weygaert}, R., \& {Jones}, B. J.~T. 2008, \mnras, 387,
  128

\bibitem[{{Pohlen} \& {Trujillo}(2006)}]{2006A&A...454..759P}
{Pohlen}, M. \& {Trujillo}, I. 2006, \aap, 454, 759

\bibitem[{{Pustilnik} \& {Tepliakova}(2011)}]{2011MNRAS.415.1188P}
{Pustilnik}, S.~A. \& {Tepliakova}, A.~L. 2011, \mnras, 415, 1188

\bibitem[{{Quilis} {et~al.}(2000){Quilis}, {Moore}, \&
  {Bower}}]{2000Sci...288.1617Q}
{Quilis}, V., {Moore}, B., \& {Bower}, R. 2000, Science, 288, 1617

\bibitem[{{Rojas} {et~al.}(2004){Rojas}, {Vogeley}, {Hoyle}, \&
  {Brinkmann}}]{2004ApJ...617...50R}
{Rojas}, R.~R., {Vogeley}, M.~S., {Hoyle}, F., \& {Brinkmann}, J. 2004, \apj,
  617, 50

\bibitem[{{Rojas} {et~al.}(2005){Rojas}, {Vogeley}, {Hoyle}, \&
  {Brinkmann}}]{2005ApJ...624..571R}
{Rojas}, R.~R., {Vogeley}, M.~S., {Hoyle}, F., \& {Brinkmann}, J. 2005, \apj,
  624, 571

\bibitem[{{Rom{\'a}n} {et~al.}(2023){Rom{\'a}n}, {Rich}, {Ahvazi}, {Sales},
  {Li}, {Golini}, {Trujillo}, {Knapen}, {Peletier}, \&
  {S{\'a}nchez-Alarc{\'o}n}}]{2023A&A...679A.157R}
{Rom{\'a}n}, J., {Rich}, R.~M., {Ahvazi}, N., {et~al.} 2023, \aap, 679, A157

\bibitem[{{Rom{\'a}n} {et~al.}(2020){Rom{\'a}n}, {Trujillo}, \&
  {Montes}}]{2020A&A...644A..42R}
{Rom{\'a}n}, J., {Trujillo}, I., \& {Montes}, M. 2020, \aap, 644, A42

\bibitem[{{Roth} {et~al.}(2005){Roth}, {Kelz}, {Fechner}, {Hahn}, {Bauer},
  {Becker}, {B{\"o}hm}, {Christensen}, {Dionies}, {Paschke}, {Popow}, {Wolter},
  {Schmoll}, {Laux}, \& {Altmann}}]{2005PASP..117..620R}
{Roth}, M.~M., {Kelz}, A., {Fechner}, T., {et~al.} 2005, \pasp, 117, 620

\bibitem[{{Ro{\v{s}}kar} {et~al.}(2008){Ro{\v{s}}kar}, {Debattista}, {Stinson},
  {Quinn}, {Kaufmann}, \& {Wadsley}}]{2008ApJ...675L..65R}
{Ro{\v{s}}kar}, R., {Debattista}, V.~P., {Stinson}, G.~S., {et~al.} 2008,
  \apjl, 675, L65

\bibitem[{{Ruiz-Lara} {et~al.}(2018{\natexlab{a}}){Ruiz-Lara}, {Beasley},
  {Falc{\'o}n-Barroso}, {Rom{\'a}n}, {Pinna}, {Brook}, {Di Cintio},
  {Mart{\'\i}n-Navarro}, {Trujillo}, \& {Vazdekis}}]{2018MNRAS.478.2034R}
{Ruiz-Lara}, T., {Beasley}, M.~A., {Falc{\'o}n-Barroso}, J., {et~al.}
  2018{\natexlab{a}}, \mnras, 478, 2034

\bibitem[{{Ruiz-Lara} {et~al.}(2018{\natexlab{b}}){Ruiz-Lara}, {Gallart},
  {Beasley}, {Monelli}, {Bernard}, {Battaglia}, {S{\'a}nchez-Bl{\'a}zquez},
  {Florido}, {P{\'e}rez}, \& {Mart{\'\i}n-Navarro}}]{2018A&A...617A..18R}
{Ruiz-Lara}, T., {Gallart}, C., {Beasley}, M., {et~al.} 2018{\natexlab{b}},
  \aap, 617, A18

\bibitem[{{Ruiz-Lara} {et~al.}(2016){Ruiz-Lara}, {P{\'e}rez}, {Florido},
  {S{\'a}nchez-Bl{\'a}zquez}, {M{\'e}ndez-Abreu}, {Lyubenova},
  {Falc{\'o}n-Barroso}, {S{\'a}nchez-Menguiano}, {S{\'a}nchez}, {Galbany},
  {Garc{\'\i}a-Benito}, {Gonz{\'a}lez Delgado}, {Husemann}, {Kehrig},
  {L{\'o}pez-S{\'a}nchez}, {Marino}, {Mast}, {Papaderos}, {van de Ven},
  {Walcher}, {Zibetti}, \& {CALIFA Team}}]{2016MNRAS.456L..35R}
{Ruiz-Lara}, T., {P{\'e}rez}, I., {Florido}, E., {et~al.} 2016, \mnras, 456,
  L35

\bibitem[{{Ruiz-Lara} {et~al.}(2015){Ruiz-Lara}, {P{\'e}rez}, {Gallart},
  {Alloin}, {Monelli}, {Koleva}, {Pompei}, {Beasley},
  {S{\'a}nchez-Bl{\'a}zquez}, {Florido}, {Aparicio}, {Fleurence}, {Hardy},
  {Hidalgo}, \& {Raimann}}]{2015A&A...583A..60R}
{Ruiz-Lara}, T., {P{\'e}rez}, I., {Gallart}, C., {et~al.} 2015, \aap, 583, A60

\bibitem[{{S{\'a}nchez} {et~al.}(2007){S{\'a}nchez}, {Aceituno}, {Thiele},
  {P{\'e}rez-Ram{\'\i}rez}, \& {Alves}}]{2007PASP..119.1186S}
{S{\'a}nchez}, S.~F., {Aceituno}, J., {Thiele}, U., {P{\'e}rez-Ram{\'\i}rez},
  D., \& {Alves}, J. 2007, \pasp, 119, 1186

\bibitem[{{S{\'a}nchez} {et~al.}(2016{\natexlab{a}}){S{\'a}nchez},
  {Garc{\'\i}a-Benito}, {Zibetti}, {Walcher}, {Husemann}, {Mendoza}, {Galbany},
  {Falc{\'o}n-Barroso}, {Mast}, {Aceituno}, {Aguerri}, {Alves}, {Amorim},
  {Ascasibar}, {Barrado-Navascues}, {Barrera-Ballesteros}, {Bekerait{\`e}},
  {Bland-Hawthorn}, {Cano D{\'\i}az}, {Cid Fernandes}, {Cavichia}, {Cortijo},
  {Dannerbauer}, {Demleitner}, {D{\'\i}az}, {Dettmar}, {de
  Lorenzo-C{\'a}ceres}, {del Olmo}, {Galazzi}, {Garc{\'\i}a-Lorenzo}, {Gil de
  Paz}, {Gonz{\'a}lez Delgado}, {Holmes}, {Igl{\'e}sias-P{\'a}ramo}, {Kehrig},
  {Kelz}, {Kennicutt}, {Kleemann}, {Lacerda}, {L{\'o}pez Fern{\'a}ndez},
  {L{\'o}pez S{\'a}nchez}, {Lyubenova}, {Marino}, {M{\'a}rquez},
  {Mendez-Abreu}, {Moll{\'a}}, {Monreal-Ibero}, {Ortega Minakata},
  {Torres-Papaqui}, {P{\'e}rez}, {Rosales-Ortega}, {Roth},
  {S{\'a}nchez-Bl{\'a}zquez}, {Schilling}, {Spekkens}, {Vale Asari}, {van den
  Bosch}, {van de Ven}, {Vilchez}, {Wild}, {Wisotzki}, {Y{\i}ld{\i}r{\i}m}, \&
  {Ziegler}}]{2016A&A...594A..36S}
{S{\'a}nchez}, S.~F., {Garc{\'\i}a-Benito}, R., {Zibetti}, S., {et~al.}
  2016{\natexlab{a}}, \aap, 594, A36

\bibitem[{{S{\'a}nchez} {et~al.}(2012){S{\'a}nchez}, {Kennicutt}, {Gil de Paz},
  {van de Ven}, {V{\'\i}lchez}, {Wisotzki}, {Walcher}, {Mast}, {Aguerri},
  {Albiol-P{\'e}rez}, {Alonso-Herrero}, {Alves}, {Bakos}, {Bart{\'a}kov{\'a}},
  {Bland-Hawthorn}, {Boselli}, {Bomans}, {Castillo-Morales}, {Cortijo-Ferrero},
  {de Lorenzo-C{\'a}ceres}, {Del Olmo}, {Dettmar}, {D{\'\i}az}, {Ellis},
  {Falc{\'o}n-Barroso}, {Flores}, {Gallazzi}, {Garc{\'\i}a-Lorenzo},
  {Gonz{\'a}lez Delgado}, {Gruel}, {Haines}, {Hao}, {Husemann},
  {Igl{\'e}sias-P{\'a}ramo}, {Jahnke}, {Johnson}, {Jungwiert}, {Kalinova},
  {Kehrig}, {Kupko}, {L{\'o}pez-S{\'a}nchez}, {Lyubenova}, {Marino},
  {M{\'a}rmol-Queralt{\'o}}, {M{\'a}rquez}, {Masegosa}, {Meidt},
  {Mendez-Abreu}, {Monreal-Ibero}, {Montijo}, {Mour{\~a}o}, {Palacios-Navarro},
  {Papaderos}, {Pasquali}, {Peletier}, {P{\'e}rez}, {P{\'e}rez}, {Quirrenbach},
  {Rela{\~n}o}, {Rosales-Ortega}, {Roth}, {Ruiz-Lara},
  {S{\'a}nchez-Bl{\'a}zquez}, {Sengupta}, {Singh}, {Stanishev}, {Trager},
  {Vazdekis}, {Viironen}, {Wild}, {Zibetti}, \&
  {Ziegler}}]{2012A&A...538A...8S}
{S{\'a}nchez}, S.~F., {Kennicutt}, R.~C., {Gil de Paz}, A., {et~al.} 2012,
  \aap, 538, A8

\bibitem[{{S{\'a}nchez} {et~al.}(2016{\natexlab{b}}){S{\'a}nchez}, {P{\'e}rez},
  {S{\'a}nchez-Bl{\'a}zquez}, {Garc{\'\i}a-Benito}, {Ibarra-Mede},
  {Gonz{\'a}lez}, {Rosales-Ortega}, {S{\'a}nchez-Menguiano}, {Ascasibar},
  {Bitsakis}, {Law}, {Cano-D{\'\i}az}, {L{\'o}pez-Cob{\'a}}, {Marino}, {Gil de
  Paz}, {L{\'o}pez-S{\'a}nchez}, {Barrera-Ballesteros}, {Galbany}, {Mast},
  {Abril-Melgarejo}, \& {Roman-Lopes}}]{2016RMxAA..52..171S}
{S{\'a}nchez}, S.~F., {P{\'e}rez}, E., {S{\'a}nchez-Bl{\'a}zquez}, P., {et~al.}
  2016{\natexlab{b}}, \rmxaa, 52, 171

\bibitem[{{S{\'a}nchez} {et~al.}(2016{\natexlab{c}}){S{\'a}nchez}, {P{\'e}rez},
  {S{\'a}nchez-Bl{\'a}zquez}, {Gonz{\'a}lez}, {Ros{\'a}les-Ortega},
  {Cano-D{\'\i}az}, {L{\'o}pez-Cob{\'a}}, {Marino}, {Gil de Paz}, {Moll{\'a}},
  {L{\'o}pez-S{\'a}nchez}, {Ascasibar}, \&
  {Barrera-Ballesteros}}]{2016RMxAA..52...21S}
{S{\'a}nchez}, S.~F., {P{\'e}rez}, E., {S{\'a}nchez-Bl{\'a}zquez}, P., {et~al.}
  2016{\natexlab{c}}, \rmxaa, 52, 21

\bibitem[{{S{\'a}nchez-Alarc{\'o}n} {et~al.}(2023){S{\'a}nchez-Alarc{\'o}n},
  {Rom{\'a}n}, {Knapen}, {Verdes-Montenegro}, {Comer{\'o}n}, {Rich}, {Beckman},
  {Argudo-Fern{\'a}ndez}, {Ram{\'\i}rez-Moreta}, {Blasco}, {Unda-Sanzana},
  {Garrido}, \& {S{\'a}nchez-Exposito}}]{2023A&A...677A.117S}
{S{\'a}nchez-Alarc{\'o}n}, P.~M., {Rom{\'a}n}, J., {Knapen}, J.~H., {et~al.}
  2023, \aap, 677, A117

\bibitem[{{S{\'a}nchez-Bl{\'a}zquez} {et~al.}(2009){S{\'a}nchez-Bl{\'a}zquez},
  {Courty}, {Gibson}, \& {Brook}}]{2009MNRAS.398..591S}
{S{\'a}nchez-Bl{\'a}zquez}, P., {Courty}, S., {Gibson}, B.~K., \& {Brook},
  C.~B. 2009, \mnras, 398, 591

\bibitem[{{S{\'a}nchez-Bl{\'a}zquez} {et~al.}(2006){S{\'a}nchez-Bl{\'a}zquez},
  {Peletier}, {Jim{\'e}nez-Vicente}, {Cardiel}, {Cenarro},
  {Falc{\'o}n-Barroso}, {Gorgas}, {Selam}, \& {Vazdekis}}]{2006MNRAS.371..703S}
{S{\'a}nchez-Bl{\'a}zquez}, P., {Peletier}, R.~F., {Jim{\'e}nez-Vicente}, J.,
  {et~al.} 2006, \mnras, 371, 703

\bibitem[{{S{\'a}nchez-Bl{\'a}zquez} {et~al.}(2014){S{\'a}nchez-Bl{\'a}zquez},
  {Rosales-Ortega}, {M{\'e}ndez-Abreu}, {P{\'e}rez}, {S{\'a}nchez}, {Zibetti},
  {Aguerri}, {Bland-Hawthorn}, {Catal{\'a}n-Torrecilla}, {Cid Fernandes}, {de
  Amorim}, {de Lorenzo-Caceres}, {Falc{\'o}n-Barroso}, {Galazzi}, {Garc{\'\i}a
  Benito}, {Gil de Paz}, {Gonz{\'a}lez Delgado}, {Husemann},
  {Iglesias-P{\'a}ramo}, {Jungwiert}, {Marino}, {M{\'a}rquez}, {Mast},
  {Mendoza}, {Moll{\'a}}, {Papaderos}, {Ruiz-Lara}, {van de Ven}, {Walcher}, \&
  {Wisotzki}}]{2014A&A...570A...6S}
{S{\'a}nchez-Bl{\'a}zquez}, P., {Rosales-Ortega}, F.~F., {M{\'e}ndez-Abreu},
  J., {et~al.} 2014, \aap, 570, A6

\bibitem[{{Sarzi} {et~al.}(2006){Sarzi}, {Falc{\'o}n-Barroso}, {Davies},
  {Bacon}, {Bureau}, {Cappellari}, {de Zeeuw}, {Emsellem}, {Fathi},
  {Krajnovi{\'c}}, {Kuntschner}, {McDermid}, \&
  {Peletier}}]{2006MNRAS.366.1151S}
{Sarzi}, M., {Falc{\'o}n-Barroso}, J., {Davies}, R.~L., {et~al.} 2006, \mnras,
  366, 1151

\bibitem[{{Scudder} {et~al.}(2012){Scudder}, {Ellison}, {Torrey}, {Patton}, \&
  {Mendel}}]{2012MNRAS.426..549S}
{Scudder}, J.~M., {Ellison}, S.~L., {Torrey}, P., {Patton}, D.~R., \& {Mendel},
  J.~T. 2012, \mnras, 426, 549

\bibitem[{{Seidel} {et~al.}(2015){Seidel}, {Falc{\'o}n-Barroso},
  {Mart{\'\i}nez-Valpuesta}, {D{\'\i}az-Garc{\'\i}a}, {Laurikainen}, {Salo}, \&
  {Knapen}}]{2015MNRAS.451..936S}
{Seidel}, M.~K., {Falc{\'o}n-Barroso}, J., {Mart{\'\i}nez-Valpuesta}, I.,
  {et~al.} 2015, \mnras, 451, 936

\bibitem[{{Simon} \& {Geha}(2007)}]{2007ApJ...670..313S}
{Simon}, J.~D. \& {Geha}, M. 2007, \apj, 670, 313

\bibitem[{{Stanonik} {et~al.}(2009){Stanonik}, {Platen}, {Arag{\'o}n-Calvo},
  {van Gorkom}, {van de Weygaert}, {van der Hulst}, \&
  {Peebles}}]{2009ApJ...696L...6S}
{Stanonik}, K., {Platen}, E., {Arag{\'o}n-Calvo}, M.~A., {et~al.} 2009, \apjl,
  696, L6

\bibitem[{{Sutter} {et~al.}(2015){Sutter}, {Lavaux}, {Hamaus}, {Pisani},
  {Wandelt}, {Warren}, {Villaescusa-Navarro}, {Zivick}, {Mao}, \&
  {Thompson}}]{2015A&C.....9....1S}
{Sutter}, P.~M., {Lavaux}, G., {Hamaus}, N., {et~al.} 2015, Astronomy and
  Computing, 9, 1

\bibitem[{{Tempel} {et~al.}(2017){Tempel}, {Tuvikene}, {Kipper}, \&
  {Libeskind}}]{2017A&A...602A.100T}
{Tempel}, E., {Tuvikene}, T., {Kipper}, R., \& {Libeskind}, N.~I. 2017, \aap,
  602, A100

\bibitem[{{van de Weygaert} {et~al.}(2011){van de Weygaert}, {Kreckel},
  {Platen}, {Beygu}, {van Gorkom}, {van der Hulst}, {Arag{\'o}n-Calvo},
  {Peebles}, {Jarrett}, {Rhee}, {Kova{\v{c}}}, \& {Yip}}]{2011ASSP...27...17V}
{van de Weygaert}, R., {Kreckel}, K., {Platen}, E., {et~al.} 2011, in
  Astrophysics and Space Science Proceedings, Vol.~27, Environment and the
  Formation of Galaxies: 30 years later, 17

\bibitem[{{van de Weygaert} \& {Platen}(2011)}]{2011IJMPS...1...41V}
{van de Weygaert}, R. \& {Platen}, E. 2011, in International Journal of Modern
  Physics Conference Series, Vol.~1, International Journal of Modern Physics
  Conference Series, 41--66

\bibitem[{{Vazdekis} {et~al.}(2015){Vazdekis}, {Coelho}, {Cassisi},
  {Ricciardelli}, {Falc{\'o}n-Barroso}, {S{\'a}nchez-Bl{\'a}zquez}, {La
  Barbera}, {Beasley}, \& {Pietrinferni}}]{2015MNRAS.449.1177V}
{Vazdekis}, A., {Coelho}, P., {Cassisi}, S., {et~al.} 2015, \mnras, 449, 1177

\bibitem[{{Vazdekis} {et~al.}(2016){Vazdekis}, {Koleva}, {Ricciardelli},
  {R{\"o}ck}, \& {Falc{\'o}n-Barroso}}]{2016MNRAS.463.3409V}
{Vazdekis}, A., {Koleva}, M., {Ricciardelli}, E., {R{\"o}ck}, B., \&
  {Falc{\'o}n-Barroso}, J. 2016, \mnras, 463, 3409

\bibitem[{{Watkins} {et~al.}(2024){Watkins}, {Kaviraj}, {Collins}, {Knapen},
  {Kelvin}, {Duc}, {Rom{\'a}n}, \& {Mihos}}]{2024MNRAS.528.4289W}
{Watkins}, A.~E., {Kaviraj}, S., {Collins}, C.~C., {et~al.} 2024, \mnras, 528,
  4289

\bibitem[{{Zernike}(1934)}]{1934MNRAS..94..377Z}
{Zernike}, F. 1934, \mnras, 94, 377

\bibitem[{{Zhang} {et~al.}(2023){Zhang}, {Martin}, {Cloutier}, {Price-Jones},
  {Abraham}, {van Dokkum}, \& {Merritt}}]{2023ApJ...948....4Z}
{Zhang}, J., {Martin}, P.~G., {Cloutier}, R., {et~al.} 2023, \apj, 948, 4

\end{thebibliography}
\begin{appendix}
\section{Additional figures}
\begin{figure*}
\centering 
\includegraphics[width = \textwidth]{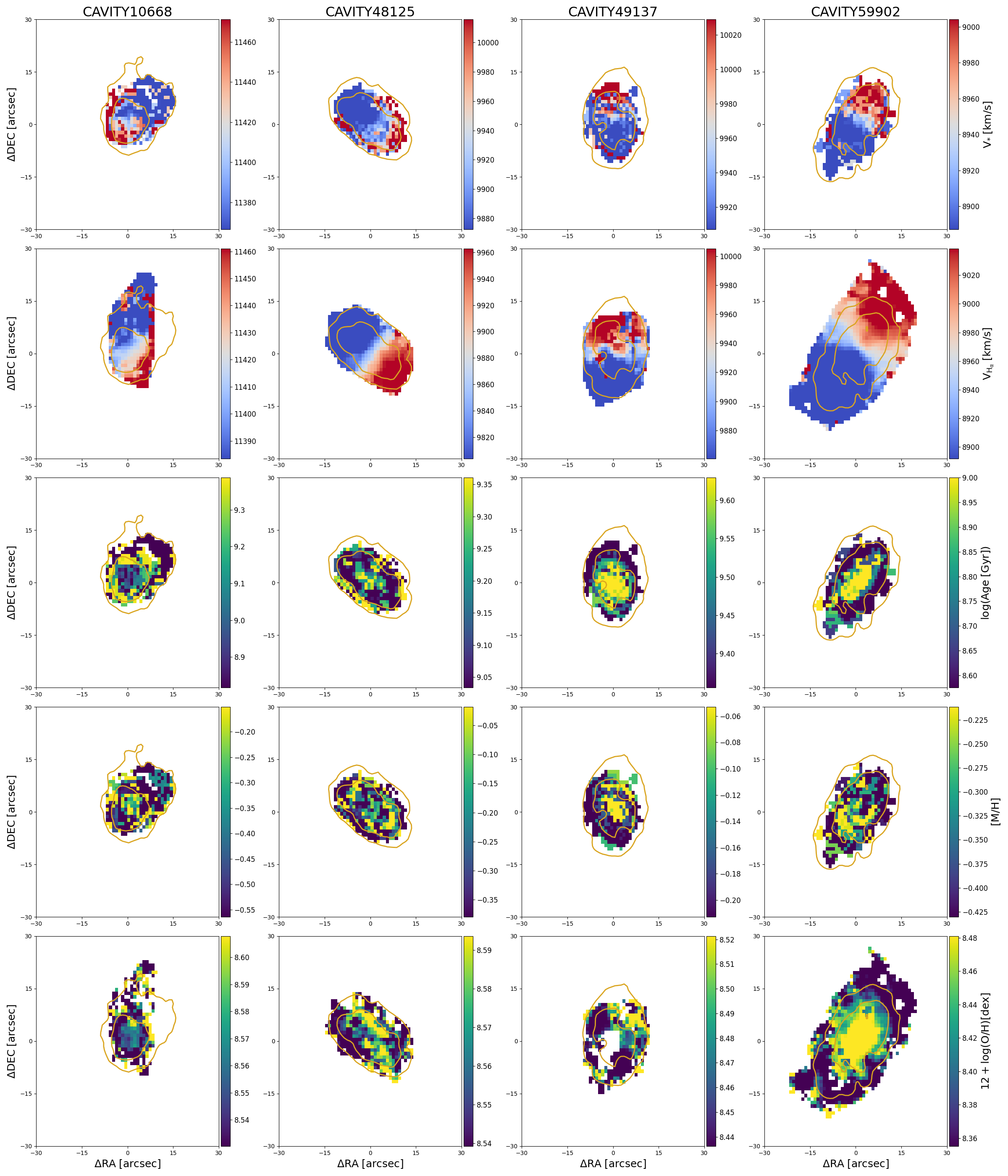} \\   
\caption{Typical science products obtained from the CAVITY IFS cubes. From left to right, we show maps for galaxies CAVITY10668, CAVITY48125, CAVITY49137, and CAVITY59902. From top to bottom, we show maps of stellar velocity, H$\alpha$ velocity, stellar age (log-scale, light-weighted), stellar metallicity ({light-weighted [Z/H]}, and gas abundance (12+log(O/H)). Contours corresponding to an S/N level of 10 and 30 are overplotted.}
\label{fig:results_ifu} 
\end{figure*}

\begin{figure*}
\centering 
\includegraphics[width = \textwidth]{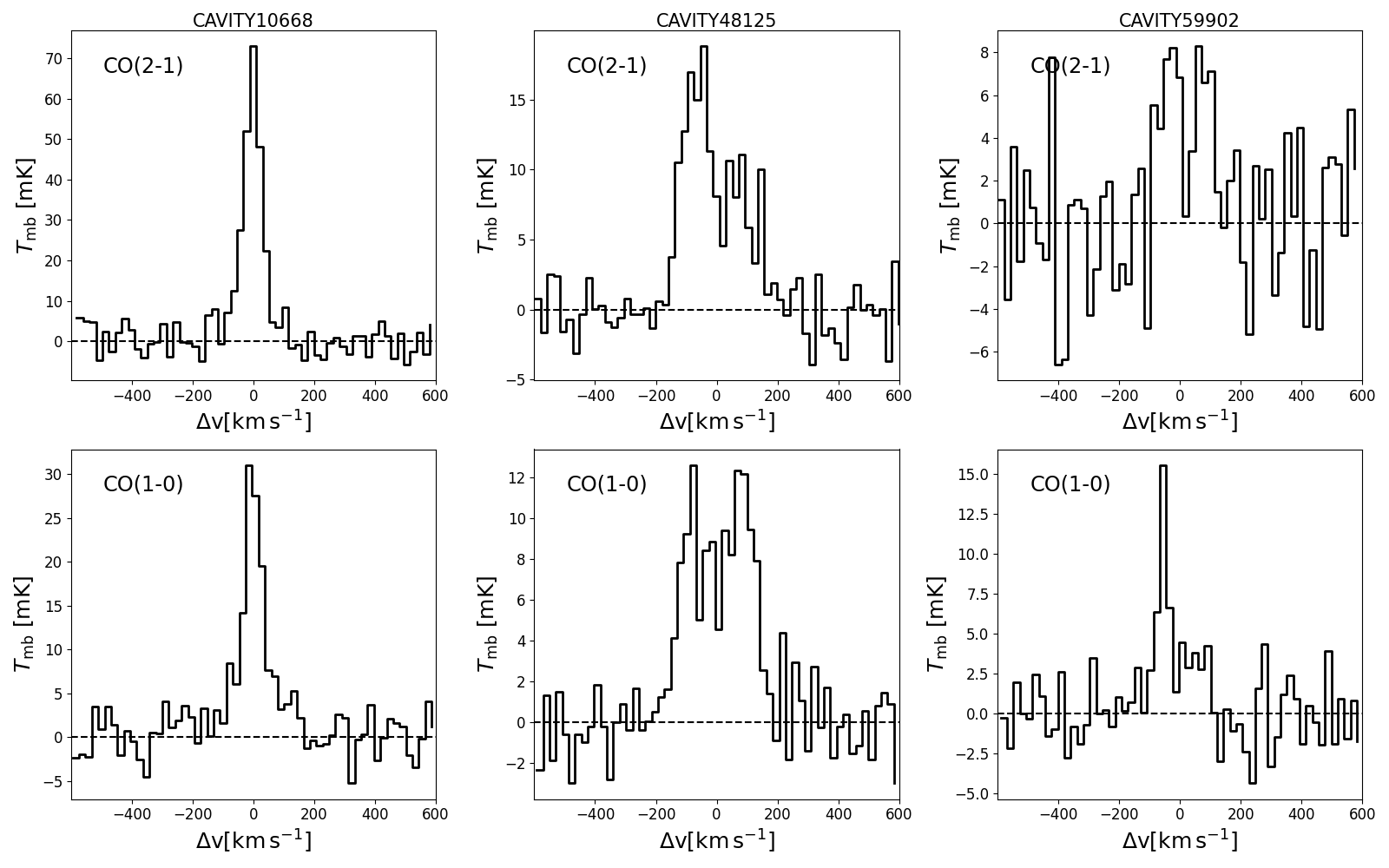} \\   
\caption{CO(2-1) (top) and CO(1-0) (bottom) spectra from our IRAM observations for CAVITY10668, CAVITY48125, and CAVITY59902 (from left to right).}
\label{fig:results_co} 
\end{figure*}

\begin{figure*}
\centering 
\includegraphics[width = \textwidth]{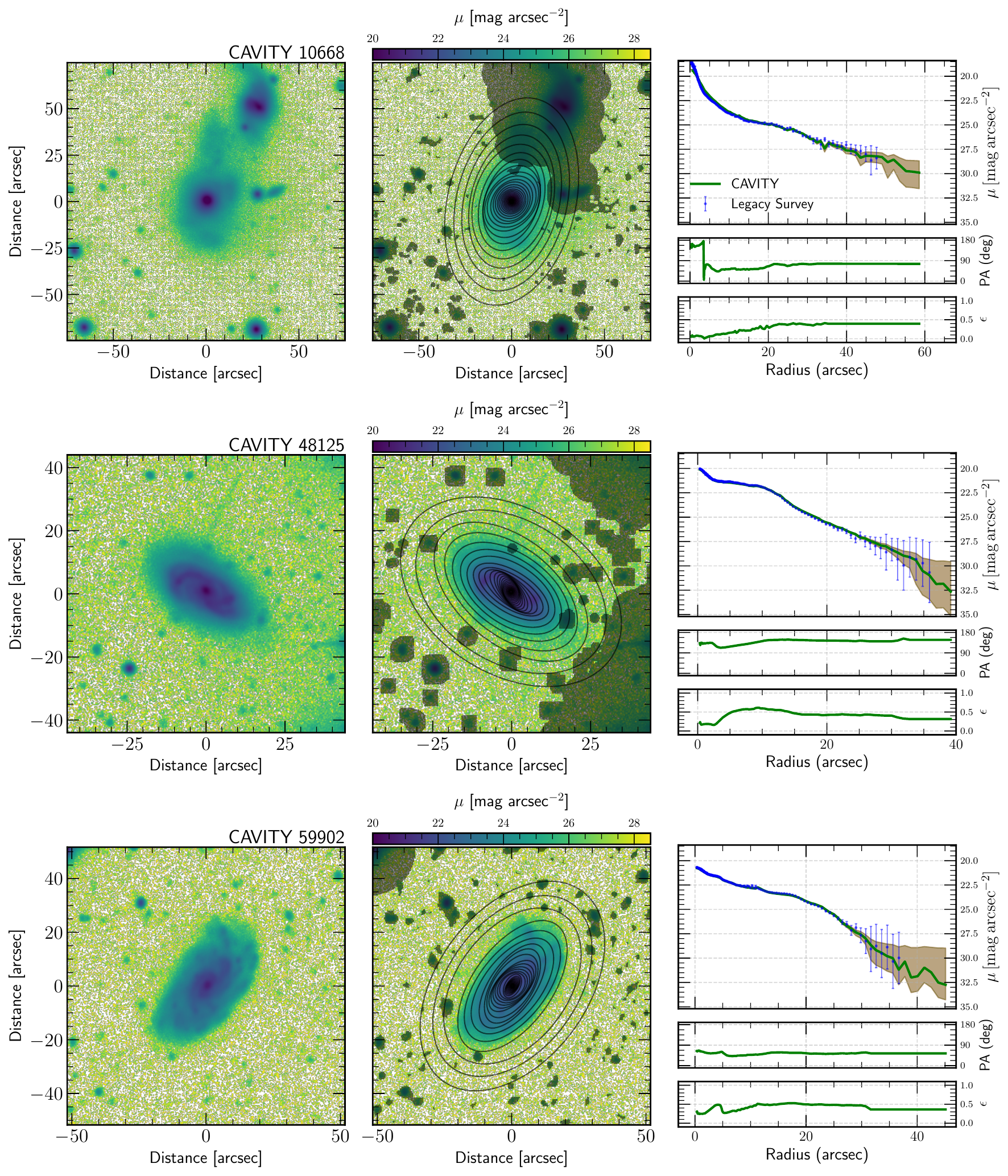} \\   
\caption{Analysis of the light distribution from the INT deep imaging campaigns for the three galaxies included in this paper with INT data so far (CAVITY10668, CAVITY48125, and CAVITY59902, from top to bottom). The left panels show the $\rm g$-band images from the INT campaign. Middle panels display, on top of the INT images, the masking scheme as well as the isophotal ellipses with varying ellipticity and position angle. The right panels show profiles of SB, position angle (PA), and ellipticity ($\epsilon$) for the three galaxies under analysis. The green lines correspond to the profiles from the INT images. The blue dots are from DECALS using the same procedure as for the INT images (see text for details).}
\label{fig:SB_profiles} 
\end{figure*}

\begin{figure*}
\centering 
\includegraphics[width = \textwidth]{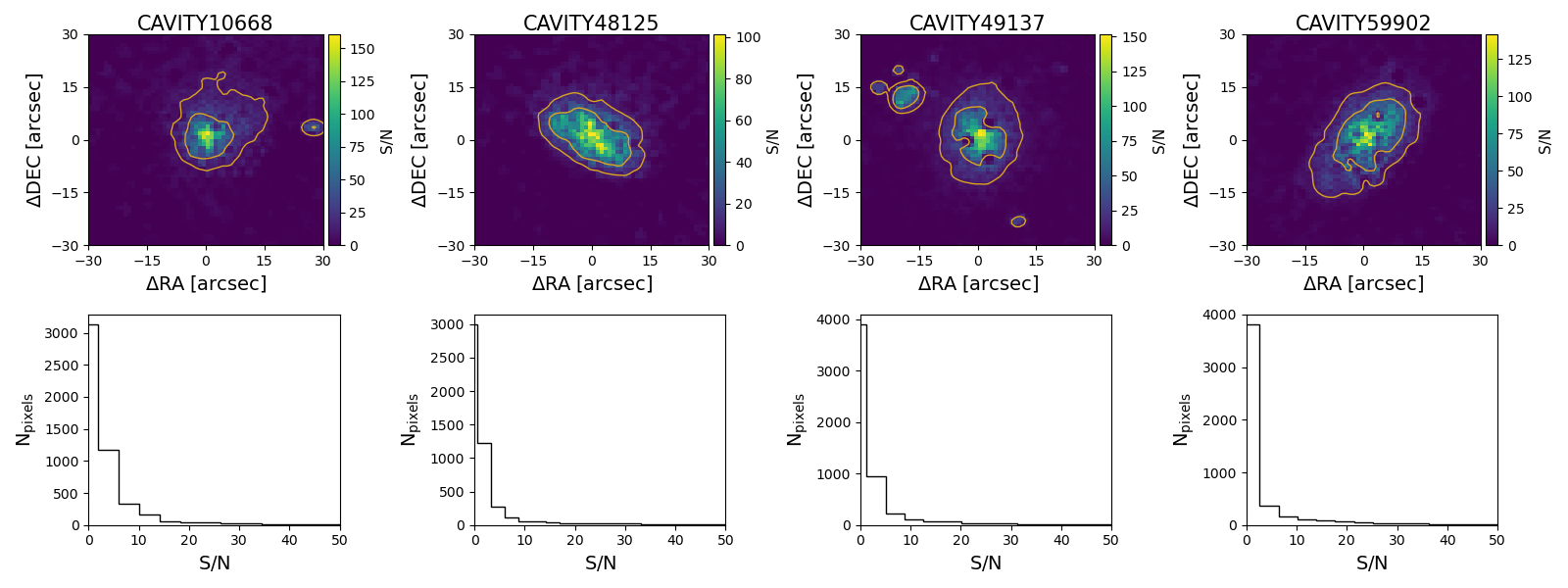} \\   
\caption{Assessment of the quality of the datacubes of the four CAVITY galaxies under analysis. Top panels: Continuum S/N maps. Contours correspond to an S/N level of 10 and 30, respectively. Bottom panels: Continuum S/N histograms. }%For aesthetic purposes we masked out all pixels with a SNR below 2.}
\label{fig:results_s2n} 
\end{figure*}

\end{appendix}
\end{document}